\documentclass[prc,aps,floats,twocolumn,twoside,preprintnumbers,
superscriptaddress,floatfix,nofootinbib,showpacs]{revtex4}
\usepackage{graphicx,epsfig,pstricks,rotating}
\usepackage{bm}

\renewcommand{\deg}{^{\circ}}

\newcommand{\eps}{\varepsilon}

\newcommand{\pipsq}[1]{\left[ #1 \right]} 

\newcommand{\eqnb}{\begin{equation}}
\newcommand{\eqne}{\end{equation}}

\newcommand{\question}[1]{}
\newcommand{\NSt}{{\mbox{\scriptsize\it NS}}}
\newcommand{\St}{{\mbox{\scriptsize\it S}}}

\def\slashchar#1{\setbox0=\hbox{$#1$}  
   \dimen0=\wd0     
   \setbox1=\hbox{/} \dimen1=\wd1  
   \ifdim\dimen0>\dimen1   
      \rlap{\hbox to \dimen0{\hfil/\hfil}} 
      #1     
   \else     
      \rlap{\hbox to \dimen1{\hfil$#1$\hfil}} 
      /      
   \fi}      %

\begin{document}

\title{$\eta$ and $\eta'$ mesons in the Dyson-Schwinger approach at 
finite temperature}
\author{D. Horvati\' c}
\email{davorh@phy.hr}
\affiliation{Physics Department, Faculty of Science,
University of Zagreb,
Bijeni\v{c}ka c. 32, Zagreb 10000, Croatia}
\author{D. Klabu\v{c}ar\footnote{Corresponding author}}
\email{klabucar@oberon.phy.hr}
\affiliation{Physics Department, Faculty of Science,
University of Zagreb, Bijeni\v{c}ka c. 32, Zagreb 10000, Croatia}
\affiliation{Senior Associate of International Centre for 
Theoretical Physics, Trieste, Italy}
\author{A. E. Radzhabov}
\email{aradzh@theor.jinr.ru}
\affiliation{
Bogoliubov  Laboratory of Theoretical Physics,
Joint Institute for Nuclear Research,
141980 Dubna,
Russia}
\date{\today}
\begin{abstract}
\noindent
We study the temperature dependence of the pseudoscalar meson
properties in a relativistic bound-state approach exhibiting the  
chiral behavior mandated by QCD. 
Concretely, we adopt the Dyson-Schwinger approach with a rank-2
separable model interaction. 
After extending the model to the strange sector and fixing  
its parameters at zero temperature, $T=0$, 
we study the $T$-dependence of the masses and decay constants  
of all ground-state mesons in the pseudoscalar nonet.  
Of chief interest are $\eta$ and $\eta^\prime$. 
The influence of the QCD axial anomaly on them 
is successfully obtained through the Witten-Veneziano  
relation at $T=0$. The same approach is then extended to $T>0$, 
using lattice QCD results for the topological susceptibility. 
The most conspicuous finding is an increase of the $\eta^\prime$ 
mass around the chiral restoration temperature $T_{\rm Ch}$, 
which would suggest a suppression of $\eta^\prime$ production in
relativistic heavy-ion collisions. 
The increase of the $\eta^\prime$ mass may also indicate that 
the extension of the Witten-Veneziano relation to finite temperatures 
becomes unreliable around and above $T_{\rm Ch}$.
Possibilities of an improved treatment are discussed. 
\end{abstract}

\maketitle

\section{Introduction and preliminaries}
\label{Preliminaries}

The experiments at RHIC advanced the empirical knowledge about 
relativistic hot 
QCD matter dramatically \cite{Muller:2006ee,Adams:2005dq} but its 
properties are so intricate, that the understanding of the state of 
matter that has been formed is still to be developed 
\cite{Adams:2005dq}.
Prior to its observation at RHIC, the hot QCD matter, usually called 
quark-gluon plasma (QGP), was pictured as 
a perturbatively interacting gas of deconfined quarks and gluons. 
It was expected to be reached above the critical temperature 
$T_c\sim 170$ MeV \cite{2decades,Hatsuda:1985eb,DeTar:1985kx,Hatsuda:1994pi}
where the rapid increase of the effective  number of active degrees of
freedom [$\eps(T)/T^4$] was found in lattice QCD simulations
\cite{Karsch:2000ps,Karsch:2003vd}.
Nowadays it is clear that such a state will be reached only at 
significantly higher temperatures than those accessible today.
For temperatures a few times higher than the critical temperature $T_c$,
the interactions and 
correlations in the hot QCD matter are still strong (e.g., see 
Refs. \cite{Shuryak:2004tx,Stephanov:2007fk}), the recent terminology 
for it being strongly coupled QGP (sQGP) \cite{Stephanov:2007fk}.

Critical assessments of hot QCD physics, e.g. Ref. \cite{Adams:2005dq},
stress the present absence of, and the need for,
a direct, compelling ``smoking gun" signal for production of a
new form of matter in RHIC collisions. The most compelling
``smoking gun'' would be a restoration of the symmetries of the 
QCD Lagrangian in hot, dense matter, notably the 
[SU$_A$(3) flavor] chiral symmetry 
and the U$_A$(1) symmetry, which are broken in the vacuum.

Among the issues pointed out as important was also the need
to clarify the role of quark-antiquark ($q\bar q$) bound states
continuing existence above the critical temperature $T_c$
\cite{Adams:2005dq}. Namely, evidence is accumulating that 
strong correlations in the form of quark-antiquark ($q\bar q$) 
bound states and resonances still exist 
\cite{Shuryak:2004tx,Blaschke:2003ut} 
in the sQGP above the critical temperature.
While in the old paradigm even deeply bound charmonium ($c\bar c$) states 
such as $J/\Psi$ and $\eta_c$ were expected to dissociate at $T \approx T_c$,
the behavior of their spectral functions as extracted by the maximum 
entropy method \cite{Asakawa:2000pv} from 
lattice QCD simulations of mesonic correlators now indicates that they 
should persist till around $2 T_c$ \cite{Datta:2002ck,Asakawa:2003re} or even 
above \cite{Shuryak:2005pp}.
There are similar indications for light-quark mesonic bound states from
lattice QCD \cite{Karsch:2002wv} and from other methods
\cite{Shuryak:2004tx,Mannarelli:2005pz}. This motivates their study in
the framework of relativistic bound-state equations, i.e.,
the Poincar\'e-covariant 
Dyson-Schwinger (DS) approach to quark-hadron physics
\cite{Roberts:2000aa,Alkofer:2000wg,Holl:2006ni,Fischer:2006ub},
which is a continuous approach complementary to 
lattice QCD studies. The DS approach is especially valuable in 
the light-quark sector, where chiral symmetry is essential.
Namely, the DS approach is unique in that it can incorporate
the correct chiral behavior of QCD, unlike other bound-state 
approaches.
This is because the crucial low-energy QCD phenomenon of
dynamical chiral symmetry breaking (DChSB) is well-understood and 
under control in DS approach
\cite{Roberts:2000aa,Alkofer:2000wg,Holl:2006ni,Fischer:2006ub}, 
at least in the consistent rainbow-ladder approximation (RLA), 
where kernel {\it Ans\" atze} of the form 
\begin{equation}
K(p)_{ef}^{hg} = - g^2 D_{\mu\nu}^{\mathrm{eff}} (p) \,
[\frac{\lambda^a}{2}\,\gamma_{\mu}]_{eg} 
[\frac{\lambda^a}{2}\,\gamma_{\nu}]_{hf}
\label{RLAkernel}
\end{equation}
(where $e,f,g,h$ schematically represent spinor, color and flavor indices)
are used for the interactions between quarks in both Eq. (\ref{DS-equation}), 
the gap equation $S^{-1} = S^{-1}_{0} - \Sigma$ for the full, dressed 
quark propagator $S$ ($S_{0}$ is the free, bare one) and the Bethe-Salpeter 
(BS) equation (\ref{BSE}) for the meson quark-antiquark bound state $\cal M$. 
That is, in RLA, the full propagator of the quark of the flavor $q$ 
characterized by the bare mass $\widetilde{m}_q$,
is given by
\begin{equation}
       S_q(p)_{ef}^{-1} = [{\rm i} \gamma \cdot p + \widetilde{m}_q]_{ef} + 
\int S_q(\ell)_{gh} K(p-\ell)_{ef}^{hg} \frac{d^4\ell}{(2\pi)^4}  \,        
\label{DS-equation}
        \end{equation}
(in the Euclidean space), 
while the bound-state vertex $\Gamma_{\cal M} = \Gamma_{q{\bar q}'}$
of the meson $\cal M$ composed of the quark $q$ and antiquark 
${\bar q}'$, is
\begin{eqnarray}
\label{BSE}
\Gamma_{q{\bar q}'}(k,{p})_{ef} =  \qquad \qquad \qquad \qquad
\qquad \qquad \qquad \qquad \quad
\\  
\int \!
[S_q(\ell+\frac{{p}}{2}) \Gamma_{q{\bar q}'}(\ell,{p})
S_{q'}(\ell-\frac{{p}}{2}) ]_{gh} K(k-\ell)_{ef}^{hg}
\frac{d^4\ell}{(2\pi)^4}~.   \nonumber
\end{eqnarray}
In the above equations, integrations are over loop momenta, 
and $D_{\mu\nu}^{\mathrm{eff}}(p)$ is an effective gluon 
propagator which at the present stage of DS studies 
\cite{Roberts:2000aa,Alkofer:2000wg,Holl:2006ni,Fischer:2006ub}
must be modeled in the non-perturbative QCD regime 
for {\it low} momenta, $p^2 \alt 1$ GeV$^2$.

In DS approach, all light pseudoscalar mesons
($\pi^{0,\pm}, K^{0,\pm}, {\bar{K^0}}, \eta$) manifest
themselves {\it both} as quark-antiquark ($q\bar q$) bound states
{\it and} (almost-)Goldstone bosons of DChSB, which also
generates the condensates and the weak decay constants of the right 
magnitude, rather independent of the small current quark masses 
$\widetilde{m}_q$, which can even be vanishing, as $\widetilde{m}_q=0$ 
corresponds to the chiral limit.  For reviews, see, e.g.,
Refs. \cite{Roberts:2000aa,Alkofer:2000wg,Holl:2006ni,Fischer:2006ub}.
Ref. \cite{Roberts:2000aa} also reviews the studies of
QCD DS equations at $T>0$, started in \cite{Bender:1996bm}.

The restoration of the chiral symmetry
should take place at the temperature $T_{\rm Ch}$, which is 
expected to be close or maybe even equal to the 
critical temperature $T_c$ (e.g., see 
Refs. \cite{Karsch:2001cy,Gattringer:2002dv,Hatta:2003ga}).  
Most scenarios expect the restoration of the $U_A(1)$ symmetry 
as the topological susceptibility melts with temperature.
Such a situation calls for a good understanding especially of the
light pseudoscalar nonet, where pions and kaons to an excellent 
approximation, and $\eta$ and $\eta'$ to a good approximation,
are eigenstates of flavor SU(3). However, for the 
neutral states with hidden strangeness,
this will change if the gluon anomaly contribution melts with
temperature.
The most interesting effects were predicted for $\eta$ and $\eta'$.
Their production rates were predicted to be enhanced 
\cite{Kapusta:1995ww,Huang:1995fc},
because their masses were expected to fall with $T$ as the 
  restoration of the $U_A(1)$ symmetry takes place.
However, for understanding such signatures, pseudoscalar meson
masses (especially $m_\eta$ and $m_{\eta'}$) at $T > 0$
must be understood better. As a contribution, the present paper
extends to finite temperatures the successful bound-state, 
DS approach of Refs.
\cite{Klabucar:1997zi,Kekez:2000aw,Kekez:2001ph,Kekez:2005ie}
to the $\eta$-$\eta'$ complex.

The paper is organized as follows. In the next section we introduce
briefly
the chosen dynamical model, but also explain generally how in the DS 
approach one can construct the mass eigenstates in the $\eta$-$\eta'$ 
complex, be it at $T=0$ or $T>0$. In the third section, the topological 
susceptibility is related to the anomalous part of the $\eta$ and $\eta'$ 
masses, and its temperature dependence fitted. 
Sec. \ref{InterplayChi-UA1symm} explains why various relationships
between chiral and U$_A$(1) symmetries would lead to different 
temperature evolutions of the $\eta$-$\eta'$ complex, 
and Sec. \ref{Scenarios} gives the $T$-dependence of 
pseudoscalar masses for various scenarios. 
We summarize in the last section.

\section{Pseudoscalar mesons at $T\geq 0$}
\label{pseudoscalarsTgeq0}

\subsection{Model and its results in non-anomalous sector}

The non-Abelian (``gluon") QCD axial anomaly is essential 
for the $\eta$--$\eta^\prime$ complex. 
Refs. \cite{Klabucar:1997zi,Kekez:2000aw,Kekez:2001ph,Kekez:2005ie}
show how to incorporate the effects of the anomaly into the DS-approach
and achieve a successful description of $\eta$ and $\eta^\prime$
(at $T=0$). They also show
\cite{Klabucar:1997zi,Kekez:2000aw,Kekez:2001ph,Kekez:2005ie}
that this works if one first achieves what is needed for the 
{\it non}-anomalous aspect of the {$\eta$--$\eta^\prime$} complex,
namely the 
good description of pions and kaons.
While the DS approach in RLA tackles this efficiently at $T=0$, 
it is technically quite difficult to extend 
solving Eqs. (\ref{RLAkernel})-(\ref{BSE})
to non-vanishing temperatures.
We thus adopt a simple model for the strong dynamics, 
the {\it separable interaction} \cite{Blaschke:2000gd}: 
\begin{equation}
g^2 D_{\mu\nu}^{\mathrm{eff}} (p-\ell) \rightarrow
\delta_{\mu\nu} D(p^2,q^2,p\cdot q) \, ,
\label{Feynmangauge}
\end{equation}
\begin{eqnarray}
D(p^2,\ell^2,p\cdot \ell)=D_0 {\cal F}_0(p^2) {\cal F}_0(\ell^2)
+ D_1 {\cal F}_1(p^2) (p\cdot \ell ) {\cal F}_1(\ell^2)~.
\label{sepAnsatz}
\end{eqnarray}
This is an interaction with the two strength parameters $D_0$, $D_1$,
and simply modeled form factors 
\begin{eqnarray}
\label{calF0}
{\cal F}_0(p^2)=\exp(-p^2/\Lambda_0^2)  \, ,
\\
{\cal F}_1(p^2)=\frac{1+\exp(-p_0^2/\Lambda_1^2)}
                     {1+\exp((p^2-p_0^2)/\Lambda_1^2)}.
\label{calF1}
\end{eqnarray}
also used in the calculations at $T\geq 0$ in Refs.
\cite{Kalinovsky:2005kx,Horvatic:2007wu,Blaschke:2007ce},
where fitting the properties of non-strange ($NS$) 
mesons, containing just $u$- and $d$-quarks, 
fixed the parameters to
\begin{eqnarray}
\Lambda_0=758 \, {\rm MeV}, \quad
\Lambda_1=961 \, {\rm MeV}, \quad
p_0=600 \, {\rm MeV},
\label{Lambda12p0}
\\
D_0\Lambda_0^2=219 \, , \qquad D_1\Lambda_0^4=40 \, ,
\label{D0D1}
\end{eqnarray}
which we also adopt here without any further re-fitting.
As in Refs. 
\cite{Blaschke:2000gd,Kalinovsky:2005kx,Horvatic:2007wu,Blaschke:2007ce},
the Matsubara formalism is used for calculations at $T>0$.
The usage of the above separable model interaction simplifies 
greatly DS equations and calculations at finite $T$, 
while yielding equivalent results on a given level of truncation
\cite{Burden:1996nh,Blaschke:2000gd}, and a similar quality
of results for pions \cite{Blaschke:2000gd,Kalinovsky:2005kx}
and now also for kaons \cite{Horvatic:2007wu} 
(and fictitious $s{\bar s}$ pseudoscalars) at $T=0$ as in Refs.
\cite{Klabucar:1997zi,Kekez:2000aw,Kekez:2001ph,Kekez:2005ie},
which employ more realistic interactions while modeling QCD.
Most of the presently pertinent results of Ref. \cite{Horvatic:2007wu}
of the separable model at $T=0$ are summarized in
Table \ref{piKssbarTable}.  In addition, 
important are also the chiral-limit value of the ${\bar q}{q}$
condensate, $\langle {\bar q}{q}\rangle_0 = \, (-217\, {\rm MeV})^3$
and the chiral-limit value of the pion decay constant,
$f_{\pi}^0 = 89 \, {\rm MeV}$. They agree well with the values 
appropriate for QCD in the chiral limit \cite{Williams:2006vv,Bernard:2006gx}.
The important aspect of the good chiral behavior is that
the pseudoscalar masses obey the Gell-Mann--Oakes--Renner type relation
\begin{equation}
M_{q\bar q'}^2 = {\rm const} \, ({\widetilde m}_q + {\widetilde m}_{q'}) \, ,
\label{M2_prop_m}
\end{equation}
as in all consistently formulated DS approaches, where
Eq. (\ref{M2_prop_m}) is an excellent approximation even 
for  realistically heavy strange quarks, 
so that the mass of the unphysical $s{\bar s}$ pseudoscalar can 
be expressed as
\begin{equation}
 M_{s\bar s}^2 = 2 M_K^2 - M_{\pi}^2 \, . 
\label{Mssbar2=2MK2-Mpi2}
\end{equation}
[E.g., the results in Table \ref{piKssbarTable}
obey Eq. (\ref{Mssbar2=2MK2-Mpi2}) up to $\frac{1}{4}$\%.]

\begin{table}[t]
\begin{center}
\begin{tabular}{|c|c|c|c|c|c|}
\hline
$\  {P} \ $   & $\ \ \  M_{P}\ \ \  $ & $\ \  M_{P}^{exp}\ \  $ & $\ \ \  f_{P}\ \ \  $ & $\ \ \  f_{P}^{exp}\ \ \  $ \\
\hline
  $\pi$     &{0.140}&{0.1396}&{0.092}&$\ 0.0924\pm 0.0003\ $ \\
\hline
  $K$       &{0.495}&{0.4937}&{0.110}&$\ 0.1130\pm 0.0010\ $ \\
\hline
  $s\bar s$ &{0.685}&        &{0.119}&                   \\
\hline
\end{tabular}
\end{center}
\caption{
Results (in GeV) on the non-anomalous pseudoscalar mesons for $T=0$:
masses $M_{P}$ and decay constants $f_{P}$ of the pseudoscalar
$q{\bar q}'$ bound states ${P} = \pi, K$ and unphysical $s\bar s$.
The corresponding constituent masses are $m_u(0)=0.398$ GeV and
$m_s(0)=0.672$ GeV, the chiral-limit condensate is
$-\langle q\bar q\rangle_0=(0.217)^3$ GeV$^3$.
All results are obtained for the bare quark masses
${\widetilde m}_{u,d} = 5.5$ MeV and ${\widetilde m}_s = 115$ MeV
and the effective interaction parameter values
(\ref{Lambda12p0}) and (\ref{D0D1}), fixed in our
Refs. \cite{Horvatic:2007wu,Blaschke:2007ce}.
These masses and decay constants are the input for
the description of the $\eta$--$\eta^\prime$ complex.
[Later, in Eq. (\ref{etaSdef}),
we will name the unphysical $s\bar s$ pseudoscalar meson
$\eta_S$, but note that the mass $M_{\eta_S}$ (\ref{MetaS}),
introduced in the {\it NS--S} mass matrix (\ref{M2_NS-S}),
includes the contribution from the gluon anomaly,
whereas $M_{s\bar s}$ does not.]
}
\label{piKssbarTable}
\end{table}

\begin{figure}[t]
\centerline{\includegraphics[width=80mm,angle=0]{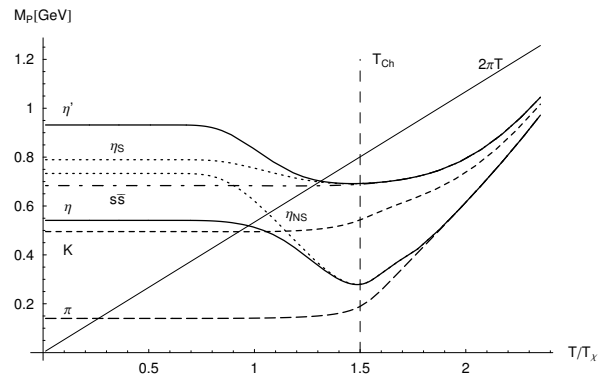}}
\caption{The dependence of the pseudoscalar meson masses on the
relative temperature $T/T_\chi$, where the temperature scale 
(characterizing $\eta$-$\eta'$ behavior) is, for 
illustrative purposes, temporarily chosen to be 
(unrealistic) $T_\chi = 2/3 \, T_{\rm Ch} = 0.667 \, T_{\rm Ch}$.
The chiral restoration temperature
$T_{\rm Ch}$ is marked by the thin dashed vertical line.
The masses which do not receive contributions from the gluon anomaly,
$M_{\pi}(T)$, $M_{K}(T)$ and $M_{s\bar s}(T)$, are
depicted by the long-dashed, short dashed and dash-dotted
curves, respectively. They practically do not change with
temperature till $T \approx 0.95 \, T_{\rm Ch}$, after which
they rise monotonically towards the thin diagonal line, which
(in all figures displaying masses)
represents twice the zeroth Matsubara frequency, $2\pi T$. This
is the limit to which meson masses should ultimately approach
from below at still higher temperatures, where $q\bar q$ states
should totally dissolve into a gas of weakly interacting quarks
and antiquarks. The masses in the $\eta$-$\eta'$ complex exhibit
different behavior due to the gluon anomaly, as will be explained 
below: the lower solid curve is $M_{\eta}(T)$, and the upper 
solid curve is $M_{\eta'}(T)$. The lower and upper dotted curves 
are, respectively, the $\eta_\NSt$ and $\eta_\St$ masses. 
 }
\label{figPseudoTc85}
\end{figure}

\begin{figure}[!tbp]
\centerline{\includegraphics[width=80mm,angle=0]{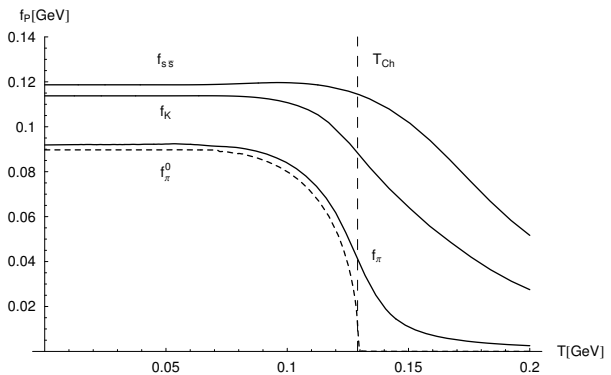}}
\caption{The temperature dependence of the pseudoscalar meson
decay constants, $f_{\pi}(T)$, $f_K(T)$ and $f_{s\bar{s}}(T)$.
We observe the typical crossover behavior. Even for the purely
nonstrange pseudoscalar, the pion, $f_{\pi}(T)$ does not fall
to zero for the realistically small but non-vanishing explicit
breaking of chiral symmetry, although the decrease of $f_{\pi}(T)$
is very strong around $T=T_{\rm Ch}$, where it would even vanish
in the chiral limit (i.e., if ${\widetilde m}_q \to 0$),
as shown by the dashed curve representing the chiral-limit
pion decay constant $f_{\pi}^0(T)$.
Increased explicit breaking of chiral symmetry due to
$s$-quarks lessens temperature dependence, as shown by
$f_K(T)$ and $f_{s\bar{s}}(T)$.
   }
\label{figfp}
\end{figure}

Solving \cite{Horvatic:2007wu} Eq. (\ref{BSE}) yields 
the masses $M_{\pi}$, $M_K$ and $M_{s\bar s}$ and the BS vertices 
$\Gamma_{q{\bar q}'}$. The meson decay constants $f_{q\bar{q}'}$ 
(e.g., $f_{u\bar{s}} = f_{K}$) can then be calculated as
\begin{eqnarray}
       f_{q\bar{q}'} \;  p_\mu &=& \, N_c \, {\rm tr_s}
                       \int  \frac{d^4\ell}{(2\pi)^4}\;\times \nonumber \\
       &\times&
\gamma_5 \gamma_\mu \; S_q(\ell+\frac{{p}}{2}) \; 
\Gamma_{q\bar{q}'}(\ell;p)\; S_{\bar{q}'}(\ell-\frac{{p}}{2})~.
\label{fqqbar}
\end{eqnarray}

These calculations were extended to $T>0$. To save space, the
temperature-dependent masses of the pseudoscalars $\pi$, $K$ 
and the $s\bar s$ are not displayed in a separate figure, but
together with the anomalous masses from the $\eta$-$\eta'$ complex,
in Fig.  \ref{figPseudoTc85} [and in Figs.
\ref{figPseudoTc107}-\ref{figNf4QCD} further below,
showing same $M_\pi(T)$, $M_K(T)$ and $M_{s\bar s}(T)$,
just for variously scaled relative temperatures.]

Fig. \ref{figfp} gives the $T$-dependences of the 
decay constants for the realistic explicit chiral symmetry
breaking, but also $f_{\pi}^0(T)$, the chiral-limit version
of $f_{\pi}(T)$. Its behavior precisely shows the value of 
the chiral symmetry restoration temperature $T_{\rm Ch}$, 
because $f_{\pi}^0(T)$ vanishes at $T=T_{\rm Ch}$,
as does the chiral condensate $\langle{\bar q}q(T)\rangle_0$.
We also confirm \cite{Kalinovsky:2005kx} the result of earlier 
DS studies at finite $T$ \cite{Maris:2000ig} that the {\it scalar}
$\sigma$-meson becomes degenerate with the pion after this
temperature $T_{\rm Ch}$; in the chiral limit its mass
$M_\sigma(T)$ in fact tends to zero as $T\to T_{\rm Ch}$ [similarly
as $f_{\pi}^0(T)$], as should be at the chiral restoration.

Besides all these correct, very advantageous behaviors, 
there is the disadvantage that in the model used in Refs. 
\cite{Horvatic:2007wu,Blaschke:2007ce} and in 
the present paper, we find the value $T_{\rm Ch}=128$ MeV,
which is significantly lower than the values (167--188 MeV) 
indicated by various lattice QCD studies \cite{Petreczky:2006tu}.
This is a well-known drawback of the rank-2 separable DS
model \cite{Blaschke:2000gd}, but its low chiral symmetry restoration 
temperature can be raised by coupling Polyakov loop variable to 
the quark sector \cite{Blaschke:2007np}.   This approach 
is currently under consideration \cite{DGMORinProgress}.  
More importantly, note that in
the present paper we do not address quantitative predictions
at some specific absolute temperature; rather, as will become
clear in detail in Sec. \ref{InterplayChi-UA1symm}, we are 
interested in the dependence on the {\it relative} temperature, 
since we will study various scenarios for 
various ratios between the chiral restoration temperature
$T_{\rm Ch}$ and the temperature scale $T_\chi$ characterizing 
the disappearance of the gluon anomaly.
For this, the present model is adequate in spite of 
relatively low $T_{\rm Ch}$.

In summary, the non-anomalous physics is all modeled 
successfully \cite{Horvatic:2007wu,Blaschke:2007ce} 
for the present purposes, and is ready to serve as input for 
constructing $\eta$ and $\eta'$.

\subsection{ The anomalous sector: {$\eta$--$\eta^\prime$} complex at $T\geq 0$}
\label{etasAtTgeq0}

The construction of physical meson states in the present model starts as 
in Refs. \cite{Klabucar:1997zi,Kekez:2000aw,Kekez:2001ph,Kekez:2005ie}.
Be it at $T=0$ or $T > 0$, solving of the BS equations yields 
quark-antiquark bound-state solutions and 
the eigenvalues of the squared masses; e.g., 
$M_{u\bar{s}}^2=M_K^2$, as the pseudoscalar $u\bar{s}$ state is simply 
the positive kaon $K^+$. By their strangeness and/or charge, 
such $q\bar q'$ states (where $q' \neq q$) are protected from mixing 
with the hidden flavor $q\bar q$ states $u\bar{u}, d\bar{d}, s\bar{s}$,
for which BS equations yields the masses 
$M_{u\bar{u}}^2,M_{d\bar{d}}^2,M_{s\bar{s}}^2$.
Although the mass matrix of the light hidden flavor pseudoscalars 
is already diagonal in the basis $u\bar{u}$-$d\bar{d}$-$s\bar{s}$,
\begin{equation}
{\hat M}^2_{NA} = 
\mbox{\rm diag} 
[\, M_{u\bar{u}}^2, \, M_{d\bar{d}}^2,\, M_{s\bar{s}}^2 \,]  
 \, ,
\label{diagM2NA}
\end{equation}
the states $u\bar{u}, d\bar{d}, s\bar{s}$ do not correspond to any 
physical particles -- certainly not to the isospin eigenstates 
$\pi^0, \eta$ and $\eta'$, the neutral flavorless  
pseudoscalars found experimentally at $T=0$. Although
$M_{u\bar{u}}^2=M_{d\bar{d}}^2=M_{\pi}^2$
{\it in the isospin limit}, $M_{u\bar{u}}^2,M_{d\bar{d}}^2,
M_{s\bar{s}}^2$ do not {\it automatically}
represent any physical masses, at least not at $T=0$,
since ${\hat M}^2_{NA}$ is only the 
{\it non-anomalous} ($NA$) part of the complete mass matrix
${\hat M}^2 = {\hat M}^2_{NA} +  {\hat M}^2_A$. 
Its {\it anomalous} part ${\hat M}^2_A$, the only
nonvanishing one in the chiral limit,
in the basis $u\bar{u}$-$d\bar{d}$-$s\bar{s}$ reads
\begin{equation}
{\hat M}^2_A = \beta
\left[ \begin{array}{ccl} 1 & 1 & 1 \\
                          1 & 1 & 1 \\
                          1 & 1 & 1
        \end{array} \right]
\begin{array}{c} {\textstyle _{\rm flavor} } \\
\vspace{-2mm} \longrightarrow \\ _{\rm breaking} \end{array} 
\, \, \beta
\left[ \begin{array}{ccl} 1 & 1 & X \\
                          1 & 1 & X \\
                          X & X & X^2
        \end{array} \right]  \, , 
\label{M2Aqq}
\end{equation}
where the first matrix neglects the influence of the SU(3) flavor 
symmetry breaking, while the second is its modification which takes 
this into account as follows. Eq. (\ref{M2Aqq}) says that due to the 
gluon anomaly, there are transitions between the states $u\bar{u}, 
d\bar{d}$ and $s\bar{s}$.
However, the amplitudes for the transitions from, and into, light 
pseudoscalar $u\bar u$ and $d\bar d$ pairs are expected to be different, 
namely larger, than those for the significantly more massive $s\bar s$.
To allow for the effects of the breaking of SU(3) flavor symmetry,
we can write 
$\langle q\bar q | {\hat M}^2_A |q' \bar q' \rangle = b_q \, b_{q'}$,
where $b_q = \sqrt{\beta}$ for $q = u, d$ and $b_q = X \sqrt{\beta}$ for 
$q = s$. 
The most widely used estimate of the flavor breaking is the ratio of
$NS$ and $S$ pseudoscalar decay constants, $X = f_\pi/f_{s\bar s}$  
\cite{Feldmann:1999uf,Kekez:2000aw,Feldmann:2002kz,Kekez:2005ie}.
Since we compute both $f_\pi$ and $f_{s\bar s}$, 
our $X$ is a predicted quantity, and not a fitting parameter.
Our present model result at $T=0$, $f_\pi/f_{s\bar s} = 0.773$,
is very close to $X_{\rm exp} \approx 0.779$ extracted
phenomenologically, i.e., from the 
mass matrix featuring experimental meson masses
-- see Refs. \cite{Kekez:2000aw,Kekez:2005ie}.

In any case, the complete mass matrix ${\hat M}^2$ is very far from 
being diagonal in this basis, since phenomenology at $T=0$ indicates 
[see $\beta_{\mathrm{fit}}$ below Eq. (\ref{eqTraces})] 
that $\beta \approx 0.3$ GeV$^2$ $\gg M_{u\bar{u}}^2,M_{d\bar{d}}^2$
(and $\beta \sim M_{s\bar{s}}^2$). 

The basis suggested for neutral $(I_3=0)$ mesons by the flavor SU(3) 
quark model, and especially by almost exact isospin symmetry, is 
the octet-singlet basis with well-defined isospin quantum numbers: the 
isovector $(I=1)$ pion $\pi^0 = (u\bar{u} - d\bar{d})/\sqrt{2}$,
and the isoscalar $(I=0)$ etas, 
$\eta_8 = (u\bar{u} + d\bar{d} - 2s\bar{s})/\sqrt{6}$,
$\eta_0 = (u\bar{u} + d\bar{d} + s\bar{s})/\sqrt{3}$.
In the case of the flavor-broken SU(3), this is the {\it effective} 
octet-singlet basis, in which the {\it complete} mass matrix
${\hat M}^2 = {\hat M}^2_{NA} +  {\hat M}^2_A$ is rather close 
to diagonal in the familiar $T=0$ case.
That is, at zero temperature and for realistic flavor breaking, 
$\eta$ is close to $\eta_8$ and $\eta'$ to $\eta_0$,
thanks to the presence of the gluon anomaly
(and in the chiral limit, the only non-vanishing mass is 
that of $\eta_0$, then equal to $3\beta$). In the DS approach, 
this is shown in detail (with all numerical values)
in Ref. \cite{Kekez:2005ie}. The present case is very 
similar, regardless the different dynamical model used. 

Therefore, the effective 
flavor-broken SU(3) states $\eta_8$ and $\eta_0$ will 
be close to the respective physical mesons $\eta$ and 
$\eta'$ as long as the gluon anomaly plays the similar
role as at $T=0$.
However, vanishing of the anomaly ($\beta = 0$) at sufficiently 
high $T$ would bring about the situation where the no-anomaly limit of 
the neutral flavorless pseudoscalar states is:
$\pi^0 = u\bar{u}, \eta = d\bar{d},
\eta' = s\bar{s}$ (where even $\eta'$ would be an
almost-Goldstone boson). This, in principle, opens the possibility 
to study the maximal isospin violation at high $T$ 
\cite{Kharzeev:1998kz,Kharzeev:2000na}, but as the effects of the
small difference between $u$ and $d$ quark masses are not important
for the present considerations, we stick to the isospin limit 
throughout the present paper. The pertinent states thus remain,
even in the high-$T$-limit, the combinations with well-defined 
isospin quantum numbers. As long as $u\bar u$ and $d\bar d$ are bound, 
$\pi^0$ is given by $(u\bar{u} - d\bar{d})/\sqrt{2}$ and is decoupled 
from mixing with the two isoscalar etas, for which the no-anomaly-limit 
states are given by the so-called {\it NS--S} basis of the $I=0$ subspace:
        \begin{eqnarray}
        \eta_\NSt
        &=&
        \frac{1}{\sqrt{2}} (u\bar{u} + d\bar{d})
  = \frac{1}{\sqrt{3}} \eta_8 + \sqrt{\frac{2}{3}} \eta_0~,
\label{etaNSdef}
        \\
        \eta_\St
        &=&
            s\bar{s}
  = - \sqrt{\frac{2}{3}} \eta_8 + \frac{1}{\sqrt{3}} \eta_0~.
\label{etaSdef}
        \end{eqnarray}
The $2\times 2$ submatrix of 
${\hat M}^2 = {\hat M}^2_{NA} +  {\hat M}^2_A$ pertinent to the masses
in the $\eta$--$\eta'$ complex, in this basis reads
\begin{equation}
\label{M2_NS-S}
       \pipsq{
                \begin{array}{ll}
      M_{\pi}^2 + 2 \beta  & \quad \sqrt{2} \beta X \\
        \,  \sqrt{2} \beta X    & M_{s\bar{s}}^2 + \beta X^2
                \end{array}
        }
\begin{array}{c} \vspace{-2mm} \longrightarrow \\ \phi \end{array}
        \pipsq{
                \begin{array}{ll}
                        M_\eta^2        & \,\,  0 \\
                     \,   0               & M_{\eta'}^2
                \end{array}
        },
\nonumber
\end{equation}
where the indicated diagonalization is achieved by
the {\it NS--S} mixing relations
\begin{equation}
\eta = \cos\phi \, \eta_\NSt
             - \sin\phi \, \eta_\St~,
\,\,\,
\eta^\prime = \sin\phi \, \eta_\NSt
             + \cos\phi \, \eta_\St~.
\label{eqno3}
\end{equation}
The {\it NS--S} mixing angle $\phi$, related to the
(equivalent) effective $\eta_8$-$\eta_0$ mixing angle $\theta$ as 
$\theta = \phi - \arctan \sqrt{2} =  \phi - 54.74^\circ$, 
is given by
\begin{equation}
\tan 2\phi
=  \frac{ 2 \, M_{\eta_{S}\eta_{NS}}^2 }{M_{\eta_S}^2-M_{\eta_{NS}}^2}
\equiv  \frac{ 2 \, \sqrt{2} \beta X}{M_{\eta_S}^2-M_{\eta_{NS}}^2} \, ,
\label{tan2phi}
\end{equation}
where, from the $\eta$-$\eta'$ mass matrix (\ref{M2_NS-S}),
\begin{eqnarray}
\label{MetaNS}
M_{\eta_{NS}}^2 = M_\pi^2 &+& 2\beta \, ,
\\
M_{\eta_S}^2 = M_{s\bar s}^2 &+& \beta X^2 
             = M_{s\bar s}^2 + \beta \, \frac{f_\pi^2}{f_{s\bar s}^2} \, .
\label{MetaS}
\end{eqnarray}
The theoretical $\eta$ and $\eta'$ mass eigenvalues are
\begin{eqnarray}
\label{Meta}
M_{\eta}^2 &=& \frac{1}{2} \left[ M_{\eta_{NS}}^2 + M_{\eta_{S}}^2
                          - \Delta_{\eta \eta'} \right] \, ,
\\
M_{\eta'}^2 &=& \frac{1}{2} \left[ M_{\eta_{NS}}^2 + M_{\eta_{S}}^2
                           + \Delta_{\eta \eta'} \right] \, ,
\label{MetaPrime}
\end{eqnarray}
where $\Delta_{\eta \eta'} \equiv \sqrt{(M_{\eta_{NS}}^2  -
M_{\eta_{S}}^2)^2 + 8 \beta^2 X^2}$.

\section{Topological susceptibility and $\eta$, $\eta^\prime$ }

In the non-anomalous pseudoscalar meson sector, we calculated 
$M_{\pi}$, $M_{s\bar s}$, $f_\pi$, $f_{s\bar s}$ and $X$ at zero and 
nonvanishing temperatures \cite{Horvatic:2007wu}.
Except for the anomalous contribution $\beta$, this is everything 
one needs for computing, at $T=0$ and $T > 0$, the mixing angle 
(\ref{tan2phi}) and our predictions for the masses in the 
$\eta$-$\eta'$ complex.
But, the anomaly effects, including $\beta$, 
obviously cannot be calculated in the present 
DS approach utilizing the ladder approximation. 
Fortunately, since the gluon anomaly is suppressed as $1/N_c$ in 
the expansion in the number of colors $N_c$, it was shown that
considering the gluon anomaly effect only at the level of mass shifts 
and neglecting its effects on the bound-state solutions is a meaningful
approximation \cite{Klabucar:1997zi,Kekez:2000aw,Kekez:2005ie}.
Thus, also in the present paper we avoid the U$_A$(1) problem 
by breaking nonet symmetry just at the level of the masses, 
adding by hand the anomaly contribution ${\hat M}^2_A$ to the 
calculated non-anomalous mass matrix ${\hat M}^2_{NA}$ 
to form the total mass matrix ${\hat M}^2$.  
{}From the traces in Eq. (\ref{M2_NS-S}),
\begin{equation}
\beta = \frac{1}{2+X^2} [ \, M_\eta^2 + M_{\eta'}^2
- M_{\pi}^2 - M_{s\bar s}^2 \, ]~,
\label{eqTraces}
\end{equation}
where $M_{\pi}$, $M_{s\bar s}$ and $X=f_\pi/f_{s\bar s}$ 
are already calculated quantities (see Table \ref{piKssbarTable}), 
whereas $M_\eta$ and $M_{\eta'}$ remain to be determined.

One may treat $\beta$ as a fitting parameter and fix it by 
requiring   
$M_\eta^2 + M_{\eta'}^2 = 
(M_\eta^2)_{exp} + (M_{\eta'}^2)_{exp} = 1.2169$ GeV$^2$ 
in Eq. (\ref{eqTraces}), 
yielding $\beta_{\mathrm{fit}} = 0.282$ GeV$^2$.
 
However, one can avoid treating $\beta$ as a new fitting parameter, 
since it can be extracted from the lattice. 
Using Eq. (\ref{Mssbar2=2MK2-Mpi2}) in Eq. (\ref{eqTraces}) 
gives the first equality in
\begin{equation}
   \beta \, (2 + X^2) = M_\eta^2 + M_{\eta'}^2 - 2 M_K^2 =
\frac{6}{f_\pi^2} \, \chi \, .
\label{WittenVenez}
\end{equation}
The second equality in Eq. (\ref{WittenVenez}) is the Witten-Veneziano
(WV) relation \cite{Witten:1979vv,Veneziano:1979ec} between the 
$\eta$, $\eta'$ and kaon masses and $\chi$, the topological susceptibility 
of the pure Yang-Mills gauge theory ($\chi = \chi_{\rm YM}$). 
 
The lattice results ($\chi_{\mathrm{latt}}$) on $\chi$ imply through Eq. (\ref{WittenVenez})
\begin{equation}
\beta_{\mathrm{latt}} = \frac{1}{2+X^2} \frac{6}{f_\pi^2} \, 
                        \chi_{\mathrm{latt}} \, .
\label{beta_latt}
\end{equation}
Taking for $\chi_{\mathrm{latt}}$ the central value of the weighted average
\begin{equation}
\chi(T=0) = (175.7 \pm 1.5 \, \rm MeV)^4 \,\, ,
\label{weightedAv}
\end{equation}
of the recent lattice results on the topological susceptibility 
\cite{Lucini:2004yh,DelDebbio:2004ns,Alles:2004vi} at $T=0$,  
gives $\beta_{\mathrm{latt}} = 0.260$ GeV$^2$.

\begin{table}[b]
\begin{center}\renewcommand{\arraystretch}{1.55}
\begin{tabular}{lllllllll}
\hline
 &$\sqrt{3\beta}$  &$M_{\eta_\NSt}$ &$M_{\eta_\St}$ &$M_{\eta_8}$ &$M_{\eta_0}$ &$\phi$ &$M_{\eta}$ &$M_{\eta'}$  \\
\hline
$\beta_{\mathrm{fit}}$ &919.5 &763.7 &798.0 &574.3 &944.2 &42.51$\deg$ &548.9 &958.5  \\
$\beta_{\mathrm{latt}}$ &884.0 &735.2 &790.0 &573.7 &914.8&40.82$\deg$ &543.1 &932.5  \\
Exp. & & & & & & &547.7 &957.8  \\
\hline
\end{tabular}
\end{center}
\caption{The $\eta$ and $\eta'$ masses and mixing angle at $T=0$, 
and comparison with experiment, for the cases 
$\beta = \beta_{\mathrm{fit}} = 0.282$ GeV$^2$
and $\beta = \beta_{\mathrm{latt}} = 0.260$ GeV$^2$. 
For the both cases, the input from  the non-anomalous sector
is the same: $X=0.773$ and $M_\pi$
and $M_{s\bar s}$ from Table \ref{piKssbarTable}. All masses 
$M_{\eta}$, $M_{\eta'}$, $M_{\eta_{NS,S,8,0}}$, and 
$\sqrt{3\beta}$ are in MeV.
The purely anomalous mass $\sqrt{3\beta}$ is the mass of 
$\eta'$ in the chiral limit for all three flavors, where 
$\eta$ is massless together with the other octet pseudoscalars. 
In the chiral limit, as well as the exact flavor SU(3) limit,
$\phi = \arctan \sqrt{2}$, i.e., $\theta = 0$, so that
$\eta' = \eta_0$, and $\eta = \eta_8$.}
\label{tab:eta-etap-mixingTzero}
\end{table}

Table \ref{tab:eta-etap-mixingTzero} summarizes our $T=0$ results
on $\eta$ and $\eta'$. The agreement with the experimental masses
is excellent for both $\beta = \beta_{\mathrm{fit}}$ and 
$\beta = \beta_{\mathrm{latt}}$. In the latter case, there is not 
even one new parameter introduced in addition to the parameters
already fixed in the non-anomalous, $\pi$ and $K$ sector.  The 
case $\beta = \beta_{\mathrm{latt}}$ is especially satisfying
because the lattice results for the topological susceptibility
$\chi$ exist also for $T > 0$ \cite{Alles:1996nm,Gattringer:2002mr},
and can be used to express $\beta(T > 0)$ and study the 
$\eta$-$\eta'$ complex also at nonvanishing temperatures,
if we assume that the WV relation continues to hold
also at $T>0$.

Various lattice results \cite{Alles:1996nm} on 
the topological susceptibility $\chi(T)$ are fitted well by 
the following expression,
\begin{equation}
\chi(T) = \chi(0)\left\{ \exp\left[ \left(\frac{T}{T_\chi}\right)^\kappa
- \kappa^{\alpha} \left( \frac{T_\chi}{T} \right)^\kappa \right] + 1 \right\}^{-\delta}~,
\label{fitAlles+alTopolSusc}
\end{equation}
depending on the {\it relative} temperature ${T}/{T_\chi}$, where $T_\chi$ is 
the characteristic ("melting") temperature where $\chi(T)$ starts 
decreasing appreciably.

\begin{figure}[t]
\centerline{\includegraphics[width=80mm,angle=0]{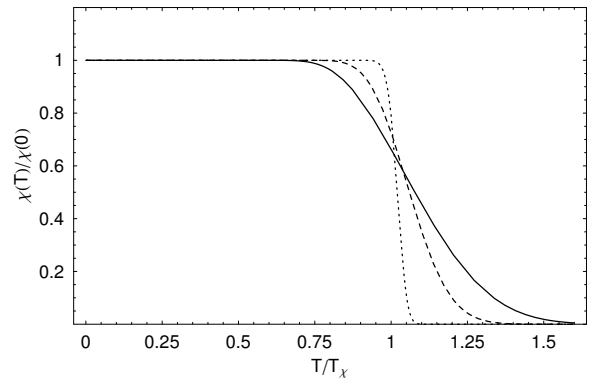}}
\caption{The relative topological susceptibility $\chi(T)/\chi(0)$ given 
by Eq. (\ref{fitAlles+alTopolSusc}) fitting the lattice data
\cite{Alles:1996nm} for the three cases: pure Yang-Mills (solid
curve, the slowest melting rate), SU(3) quenched (dashed curve, 
intermediate melting rate), and four-flavor QCD (dotted curve,
the fastest melting rate). 
}
\label{figFit3AllesTopSusc}
\end{figure}

For $\kappa=4.2$, $\alpha=0.2$ and $\delta=0.76$ (solid curve in Fig. 
\ref{figFit3AllesTopSusc}), Eq. (\ref{fitAlles+alTopolSusc}) fits well 
the lattice data \cite{Alles:1996nm} on the pure Yang-Mills 
topological susceptibility, $\chi(T) = \chi(T)_{\rm YM}$,  
which is the one consistent to use in the WV relation. However, we will 
also be interested in fitting the shape of $\chi(T)/\chi(0)$ for the 
SU(3) quenched QCD data [fit obtained with $\kappa=5.1$, $\alpha=0.5$ and 
$\delta=\kappa/4$ in Eq. (\ref{fitAlles+alTopolSusc}), dashed curve
in Fig. \ref{figFit3AllesTopSusc}], as well as that shape for the 
case of the four-flavor QCD lattice data \cite{Alles:1996nm}
(dotted curve in Fig. \ref{figFit3AllesTopSusc}, obtained with 
$\kappa=13$, $\alpha=0.5$ and $\delta=2.4$).

One extreme behavior of the topological susceptibility, 
qualitatively described by Eq. (\ref{fitAlles+alTopolSusc}) 
for low values of $T_\chi$, 
is the most traditional scenario, due to Pisarski and 
Wilczek \cite{Pisarski:1983ms} and studied, e.g., by 
Ref. \cite{Huang:1995fc}.  
Now, however, it appears to be disfavored, as then 
$\chi(T)$ melts away with $T$ rather quickly, since one supposes 
that $\chi(T)$ is due to the instanton contribution which decreases 
exponentially as the so-called ``Pisarski-Yaffe suppression factor'' 
\cite{Pisarski:1980md}. Then, $T_\chi$ would be significantly 
below $T_{\rm Ch}$, 
which may be judged too unrealistic to consider \cite{Shuryak:1993ee}.
Still, we consider it for illustrative purposes, since in the present 
bound-state approach in conjunction with $\chi(T) = \chi(T)_{\rm YM}$,
one can get falling masses and enhanced $\eta$ and $\eta'$ production 
(predicted by, e.g., Refs. \cite{Kapusta:1995ww,Huang:1995fc})
only in this case, and for $T_\chi = T_{\rm Ch}$ but with
$\chi(T)$ falling much more sharply than $\chi(T)_{\rm YM}$,
as will be shown below.  

The opposite extreme possibility is $\chi(T)\approx const$; i.e.,
U$_A$(1) symmetry is not restored around critical temperature, but 
at much higher, possibly infinite $T$, e.g., see Ref. \cite{Kogut:1998rh}. 
This possibility is approached
by Eq. (\ref{fitAlles+alTopolSusc}) for growing $T_\chi$.

At zero temperature, $\eta_\NSt$ and $\eta_\St$ are almost as 
``misaligned" with $\eta$ and $\eta'$ as they can be, as shown 
by our {\it NS--S} mixing angle close to $\phi \approx 45\deg$.
This must be so for any successful model, since it is mandated 
by phenomenology at $T=0$, as seen, e.g., in Ref. \cite{Kekez:2005ie}.
But, if the topological susceptibility vanishes when $T$ becomes 
large enough, $\beta \to 0$ causes $\phi \to 0$ in the end, along with
$\eta \to \eta_\NSt, \eta'\to \eta_\St$
and $M_\eta \to M_{\pi}(T)$, $M_{\eta'} \to M_{s\bar{s}}(T)$.
Depending on how the mass contribution of the gluon anomaly
varies with growing $T$, various interesting effects can occur
before this limit is reached.

\section{Chiral vs. U$_A$(1) 
symmetry restoration at $T > 0$}
\label{InterplayChi-UA1symm}

The temperature evolution of the $\eta$-$\eta'$ mass matrix 
(\ref{M2_NS-S}) results from the $T$-dependence not only of 
$\chi$, but also of $f_{\pi}$, $f_{s\bar{s}}$,
$M_{\pi}$ and $M_{s\bar{s}}$. As shown by  
Eq. (\ref{WittenVenez}), the ratio $\chi(T)/f_{\pi}(T)^2$ 
is crucial for the $T$-dependence of the anomalous mass contribution. 

Thus, contrary to the assumptions often made in the literature 
(e.g., Ref. \cite{Schaffner-Bielich:1999uj}),  
the decrease of the topological susceptibility does not 
imply {\it automatic} restoration of U$_A$(1) symmetry 
(e.g., see Ref. \cite{Fukushima:2001hr}). 
We will see that the assumption that it does, is correct only 
if $\chi$ falls sufficiently faster than $f_{\pi}^2$ does, so 
that $\chi(T)/f_{\pi}^2(T)$ falls towards zero as $T$ grows. 
We should thus explore what happens for various possible 
relationships between $T_\chi$ and $T_{\rm Ch}$,
the respective temperatures of the onset of the 
fast decrease of $\chi(T)$ and of $f_{\pi}(T)$.

Many papers addressing the relationship 
of restoration of the chiral vs. U$_A$(1) symmetry, e.g., Refs. 
\cite{Huang:1995fc,Fukushima:2001hr,Schaffner-Bielich:1999uj},
consider only one value of the temperature characterizing 
the melting of $\chi(T)$. 
Those of them \cite{Fukushima:2001hr,Schaffner-Bielich:1999uj} 
using the lattice results of Ref. \cite{Alles:1996nm} on the 
$T$-dependence of $\chi$,
do not employ the rather high susceptibility melting temperature (260 MeV) 
of that reference. They adopt the view of Ref. \cite{Alles:1996nm} that
this temperature ($T_\chi$ in {\it our} notation) is the critical temperature 
$T_c$, but {\it rescale} the lattice results \cite{Alles:1996nm} for 
$\chi(T)$ to a lower value, $T_c=T_\chi=150$ MeV. 
(This in the ballpark appropriate to the full QCD, but 
significantly lower than the pure Yang-Mills case, where $\chi$
consistent with the WV relation is computed.)
Rescaling of $T_\chi$ is also relevant for us, since we are 
interested in the dependence on  {\it relative} temperature, i.e.,
we study various relationships between $T_\chi$ and $T_{\rm Ch}$.
Below, we exhibit the case $T_\chi = T_{\rm Ch}$ as well as the cases 
when $T_\chi$ is roughly 15\% and 30\% below 
and above $T_{\rm Ch}$.

Recall that the present DS model gave us $T_{\rm Ch}=128$ MeV 
for the chiral restoration temperature.
This is lower than lattice results, since lattice finds that 
$T_{\rm Ch}$ seems to coincide with the critical temperature $T_c$
\cite{Karsch:2001cy,Gattringer:2002dv,Hatta:2003ga} 
and presently estimates $T_c$ to be between 167 and 188 MeV for 
2+1 flavor QCD \cite{Petreczky:2006tu}.
Nevertheless, our results have a more generic meaning than
working in this simple model would indicate at the first glance; 
using a more realistic interaction in DS approach, such as in 
Ref. \cite{Maris:2000ig}, gives $T_{\rm Ch}$ above 150 MeV, but the
results obtained there otherwise look similar to the corresponding 
results here, just rescaled to the higher $T_{\rm Ch}$.  Notably,
Ref. \cite{Maris:2000ig} also found that
$f_\pi(T)$ and $M_\sigma(T)$, vanishing at $T=T_{\rm Ch}$ 
in the chiral limit, do not vanish any more when the explicit breaking 
of chiral symmetry is introduced, but exhibit a crossover, just as 
we find in the present model.
Also, the pseudoscalar meson mass exhibits very little variation 
below $T_{\rm Ch}$ \cite{Maris:2000ig}, just as the $\pi, K$ and 
$s\bar s$ masses here. 
The main change is thus pushing the fall-off behavior of $f_\pi(T)$, 
which happens around $T_{\rm Ch}$, to higher temperatures. 
The temperature-induced changes in the $\eta$-$\eta'$ complex are 
thus given essentially by $2\beta(T) \sim 4 \chi(T)/f_{\pi}^2(T)$ 
and $\beta(T) {f_\pi^2(T)}/{f_{s\bar s}^2(T)} \sim 
                                  2 \chi(T)/{f_{s\bar s}^2(T)}$
[see Eqs. (\ref{MetaNS}) and (\ref{MetaS})].
Fig. \ref{figfp} shows that unlike $f_{\pi}(T)$, 
$f_{s\bar s}(T)$ does not decrease by an order of magnitude 
across the examined $T$-interval. Thus, the question what
decreases earlier, $\chi(T)$ or $f_{\pi}^2(T)$, 
and the related relationship between $T_\chi$ and $T_{\rm Ch}$,
is what causes important, qualitative differences in $T$-evolution 
scenarios, and not what we happen to chose for our concrete dynamical 
model in the DS approach.

Let us here again note the advantage of the DS approach for analyzing 
the hot QCD matter in which $q\bar q$ bound states and/or resonances
persist beyond the critical temperature marking not a real phase 
transition, but a relatively smooth crossover \cite{Stephanov:2007fk}.
The DS approach enables the meaningful computation of quantities 
characterizing such bound states or resonances. 
For example, note that 
Fukushima et al. \cite{Fukushima:2001hr}, who were the first 
to study the interplay of chiral symmetry restoration
and melting of $\chi$ in WV relation (\ref{WittenVenez}), uses the 
Nambu--Jona-Lasinio (NJL) model (following, e.g., the pioneering 
papers \cite{Hatsuda:1985eb,Hatsuda:1994pi}). However, the lack of 
confinement in the NJL model also leads to the disadvantage, 
important in the present context, that $\eta'$ is not
bound at all, not even at $T=0$. It decays into quarks and 
antiquarks due to its large mass, so that $M_{\eta'}$ in the 
NJL model is not a well defined quantity \cite{Fukushima:2001hr}.
On the other hand, the present DS approach does not have such
problems.   

Another advantage of the DS approach is the behavior of $f_{\pi}(T)$. 
This approach has the correct chiral behavior also for finite $T$,
so that $f_{\pi}(T)$ falls markedly at $T=T_{\rm Ch}$,  
but since the
DS approach naturally incorporates also the effects of realistic 
explicit chiral symmetry breaking, $f_{\pi}(T)$ stays non-zero 
\cite{Maris:2000ig,Roberts:2000aa}, 
albeit small, well beyond $T=T_{\rm Ch}$. This is in contrast to the 
chiral-limit $f_{\pi}^0(T)$, where $f_{\pi}^0(T \geq T_{\rm Ch})=0$ 
precludes its usage in WV relation for $T \geq T_{\rm Ch}$ 
(e.g., Ref. \cite{Fukushima:2001hr} or the studies 
\cite{Huang:1995fc,Schaffner-Bielich:1999uj,Meyer-Ortmanns:1994nt} 
employing chiral Lagrangians).

\section{Results on {$\eta$--$\eta^\prime$} complex at $T > 0$}
\label{Scenarios} 

\subsection{Yang-Mills topological susceptibility}
\label{YMtopSusc}

First, we explore the temperature dependence of the {$\eta$--$\eta^\prime$} 
complex following from the usage of the pure Yang-Mills topological 
susceptibility, $\chi(T) = \chi(T)_{\rm YM}$ since it is the 
one consistent  
with the WV relation. 
One should keep in mind that the pure Yang-Mills case 
has the slowest $T$-dependence, i.e., a significantly slower fall-off 
than cases when quarks are present (see Fig. \ref{figFit3AllesTopSusc}), 
which will motivate considering other cases in the subsection 
\ref{caveatsPossib}. 

For illustrative purposes, let us begin with an unrealistically 
low $\chi$-melting temperature, say $T_\chi = 2/3 \, T_{\rm Ch}$.
This is a representative case when the topological susceptibility
$\chi(T)$ becomes very small well before $T_{\rm Ch}$, i.e., well before
$f_{\pi}(T)$ can appreciably decrease. The temperature dependence
of $\eta$ and $\eta'$ masses for this case is shown in
Fig. \ref{figPseudoTc85}.
They remain practically as at $T=0$,
until $T\approx 0.9T_\chi$, after which the gluon anomaly 
contributions start melting, and both $M_{\eta}(T)$ and $M_{\eta'}(T)$
start falling. After $T \approx 1.2 \, T_\chi = 0.8\, T_{\rm Ch}$,
we have practically $M_{\eta}(T) = M_{\eta_\NSt}(T)$ 
and $M_{\eta'}(T) = M_{\eta_\St}(T)$
as $\eta$ becomes pure $\eta_\NSt$ and $\eta'$ becomes pure $\eta_\St$,
and even pure non-anomalous $s\bar s$ after $T \approx 1.3 \, T_\chi$. 
After $T=T_{\rm Ch}$, all masses rise, and $\eta$ soon becomes 
degenerate with the pion.

\begin{figure}[b]
\centerline{\includegraphics[width=80mm,angle=0]{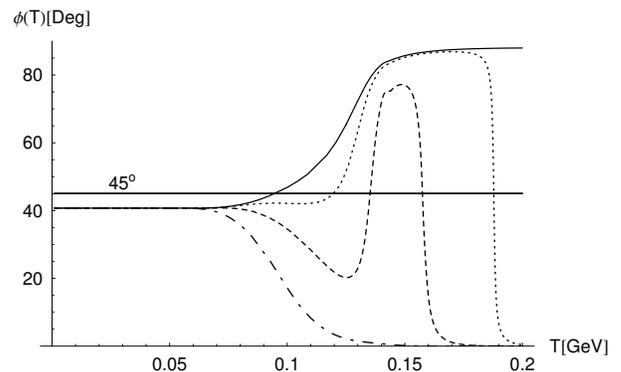}}
\caption{The temperature dependence of the $\NSt$-$\St$ mixing 
angle $\phi$ for $T_\chi = 2/3 \, T_{\rm Ch}$ (dash-dotted curve),
$T_\chi = 0.758 \, T_{\rm Ch}$ (dashed curve, the case not 
displayed in other figures), 
$T_\chi = 0.836 T_{\rm Ch}$ (dotted curve), and
$T_\chi = T_{\rm Ch}$ (solid curve). In all the cases, 
the pure Yang-Mills topological susceptibility is used 
in the WV relation (\ref{WittenVenez}).}
\label{figPHI}
\end{figure}

This behavior is reflected in the mixing angle $\phi(T)$
which for the choice $T_\chi = 2/3 \, T_{\rm Ch}$ just follows
the topological susceptibility (\ref{fitAlles+alTopolSusc}), 
falling monotonically from its $T=0$ value to zero - 
see Fig.  \ref{figPHI}.

An important, potentially experimentally recognizable feature
in Fig. \ref{figPseudoTc85} is the decrease of the masses of
$\eta'$, and especially of $\eta$. It starts already well before 
$T=T_{\rm Ch}$, where the $\eta$ mass is less than a half of its
$T=0$ value. The decrease of the masses should result in 
an increase of their (especially $\eta$'s) multiplicities.
This case, with $T_\chi$ significantly lower than $T_{\rm Ch}$, thus
illustrates well the $T$-dependences of the masses of pseudoscalars in the 
Pisarski-Wilczek scenario \cite{Pisarski:1983ms,Kapusta:1995ww,Huang:1995fc}.  

\begin{figure}[t]
\centerline{\includegraphics[width=80mm,angle=0]{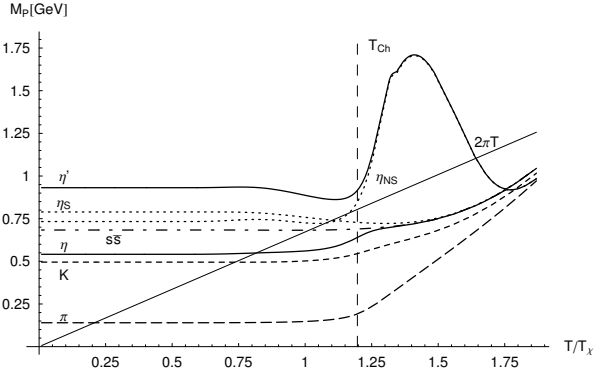}}
\caption{The relative temperature dependence, on $T/T_\chi$,
of the pseudoscalar meson masses for $T_\chi = 0.836 \, T_{\rm Ch}$,
i.e, $T_{\rm Ch} = 1.20\, T_\chi$, marked by the thin dashed vertical
line. The long-dashed, short dashed and dash-dotted curves
again represent, respectively,
$M_{\pi}(T)$, $M_{K}(T)$ and $M_{s\bar s}(T)$.
The lower and upper solid curves are again $M_{\eta}(T)$
and $M_{\eta'}(T)$, respectively, and the lower and upper
(except on a small segment) dotted
curves are $M_{\eta_\NSt}(T)$ and $M_{\eta_\St}(T)$.
  }
\label{figPseudoTc107}
\end{figure}

Let us see how the behavior changes when $T_\chi$ is increased by,
e.g., some 25\%, to $T_\chi = 0.836\, T_{\rm Ch}$. 
Unlike in Fig. \ref{figPseudoTc85}, $T=T_{\rm Ch}$ does not 
mark the onset of the approach to the asymptotic situation
[$M_{\eta'}(T)\to M_{s\bar s}(T)$, $M_{\eta}(T)\to M_\pi(T)$].     
Now, in 
Fig. \ref{figPseudoTc107}, $T=T_{\rm Ch}$ marks the strong increase 
of the $\eta'$ mass, which almost doubles by $T/T_\chi \approx 1.4$,
where it reaches its maximum and starts falling till its 
second, very narrow anti-crossing 
with the $\eta$ mass, which just grows tamely and (almost) 
monotonically. [The crossings of $\eta_{\NSt}$ and 
$\eta_{\St}$ correspond to the anti-crossings of the $\eta$ and
$\eta'$ masses (\ref{Meta}) and (\ref{MetaPrime}).] 
The increase of $T_\chi$ also pushed 
the onset of the approach to the asymptotic situation 
beyond this second anti-crossing.

This all happens because now the
decrease of the pion decay constant $f_{\pi}(T)$ starts sufficiently
before the decrease of the topological susceptibility, so that a large
amplification of $\beta(T) \sim \chi(T)/f_{\pi}^2(T)$ occurs, and over
a significant temperature interval. In contrast,
$\beta(T) X(T)^2 \sim \chi(T)/f_{s\bar s}^2(T)$ is not so 
significantly enhanced.
Eqs. (\ref{MetaNS}), (\ref{MetaS}) and (\ref{WittenVenez}) 
show why $M_{\eta_\NSt}$ is larger than $M_{\eta_\St}$ 
in the interval $1.1\, T_\chi \alt T \alt 1.75\, T_\chi$
and is the largest contributor to $M_{\eta'}$.
In the $T$-interval where $M_{\eta_\NSt}(T)$ is significantly
larger than other masses and $\beta(T) X(T)^2$, expansion of
$\Delta_{\eta \eta'}$ 
in Eq. (\ref{Meta}) shows that the $\eta$ mass is approximately the
non-anomalous $s\bar s$ mass $M_{s\bar s}(T)$, with the
gluon-anomaly contributions suppressed by $1/M_{\eta_{NS}}(T)$:
\begin{equation}
M_{\eta}^2 \approx M_{s\bar s}^2 -
\frac{1}{2} \left[\frac{M_{s\bar s}^2}{M_{\eta_{NS}}} +
\frac{\beta X^2}{M_{\eta_{NS}}} \right]^2 + \,  ... \, \, .
\end{equation}
In this temperature interval, the heavier particle, $\eta'$,
is predominantly non-strange, $\eta_{NS}$, while the lighter one,
$\eta$, is mostly strange, $\eta_{S}$. 
Eq. (\ref{tan2phi}) shows that as 
$M_{\eta_\NSt}$ grows, so does $\phi(T)$. It surpasses $45\deg$
when $M_{\eta_\NSt}$ surpasses $M_{\eta_\St}$ -- compare Figs.
\ref{figPHI} and \ref{figPseudoTc107}.
In fact, after $T\approx 1.25\, T_\chi = T_{\rm Ch}$, it climbs 
close to $\phi(T)\sim 90\deg$, as $\eta' \sim \eta_{NS}$ 
and $\eta \sim -\eta_S$.
After, as $\chi(T)\to 0$, $M_{\eta_{NS}}(T)$ falls back towards 
$M_{\pi}(T)$, and $\phi(T)$ falls [through $45\deg$ where
again $M_{\eta_{NS}} = M_{\eta_S}$, while $M_{\eta'}(T)$ 
and $M_{\eta}(T)$ anti-cross] towards zero,
and etas quickly reach their no-anomaly limit
(\ref{etaNSdef})-(\ref{etaSdef}): $\eta = \eta_{NS}$
degenerate with the pion, and $\eta' = \eta_S$ 
with the non-anomalous mass $M_{s\bar s}(T)$.

\begin{figure}[b]
\centerline{\includegraphics[width=80mm,angle=0]{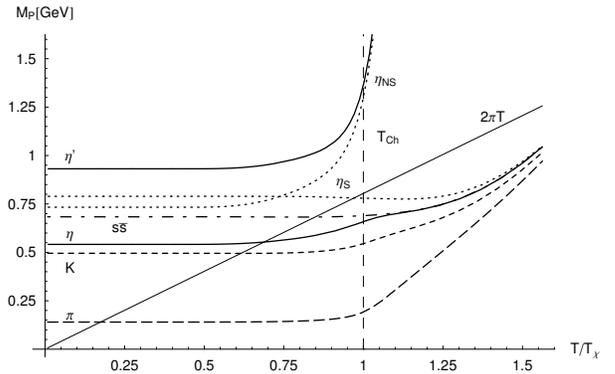}}
\caption{The relative temperature dependence, on $T/T_\chi$,
of the pseudoscalar meson masses for $T_\chi = T_{\rm Ch}$,
marked by the thin dashed vertical line. The meaning of
all curves, lines and other symbols is the same as in the
Figures \ref{figPseudoTc85} and \ref{figPseudoTc107}.
The most conspicuous is the increase of $M_{\eta'}(T)$,
which become as high as 5 GeV around $T=1.3\, T_\chi$.
  }
\label{figPseudoTc128}
\end{figure}

The crucial role of the chiral restoration temperature 
$T_{\rm Ch}$ is already obvious. We assume next that the 
topological-susceptibility-melting temperature $T_\chi$ is still
higher and equal to this temperature, i.e., $T_\chi = T_{\rm Ch}$. 
This case is illustrated in Fig. \ref{figPseudoTc128}. It exhibits
several interesting features. The most obvious one is such a strong
increase of the $\eta'$ mass that soon after $T = T_{\rm Ch}$, it exceeds
the scale of the figure. Of course, it does not indicate any divergence;
after reaching the maximum value of around 5 GeV, $M_{\eta'}(T)$ falls down.
It is just that the decreasing part of the $M_{\eta'}(T)$ curve below 1.6 GeV
and the close $\eta$-$\eta'$ anti-crossing are pushed outside the examined
temperature interval by widening of the interval where $M_{\eta_{NS}}(T)$
is enhanced. This results in the mixing angle $\phi(T)$ which
in the displayed $T$-interval ($T \leq 200$ MeV), does not return down
from the value of approximately $90\deg$; i.e., $\eta'$ remains
almost pure $\eta_{NS}$, and $\eta$ almost pure $\eta_S$ in the
displayed $T$-interval. The fall of $M_{\eta'}$ and approach
to the no-anomaly asymptotic situation is thus postponed to higher
temperatures now that we increased $T_\chi$, the `melting temperature' 
of the (anyway comparatively slowly decreasing) Yang-Mills $\chi$.
On the other hand, since the decrease of $f_\pi(T)$ is determined
by $T_{\rm Ch}$, increasing $T_\chi$
lowers the value of the relative temperature 
where $M_{\eta_{NS}}(T)$ begins its fast climb.

All these features are readily understood on the basis of the detailed 
explanation of the previous case $T_\chi = 0.836 \, T_{\rm Ch}$, but 
they are much more pronounced with respect to this case with just
16\% lower $T_\chi$. 
The described trends continue for $T_\chi > T_{\rm Ch}$.
We illustrate this through the 
choice $T_\chi = 1.17 \, T_{\rm Ch}$, which in our model
corresponds to $T_\chi = 150$ MeV.
This value is equal to $T_\chi$ and the critical temperature
assumed, e.g., in the typical examples
\cite{Huang:1995fc,Fukushima:2001hr,Schaffner-Bielich:1999uj}
of studies paralleling ours. These assumptions are obviously
motivated by 150 MeV being roughly the value of the critical,
transition temperature $T_c$ found on the lattice for
3-flavor QCD \cite{Karsch:2001cy}.

\begin{figure}[t]
\centerline{\includegraphics[width=80mm,angle=0]{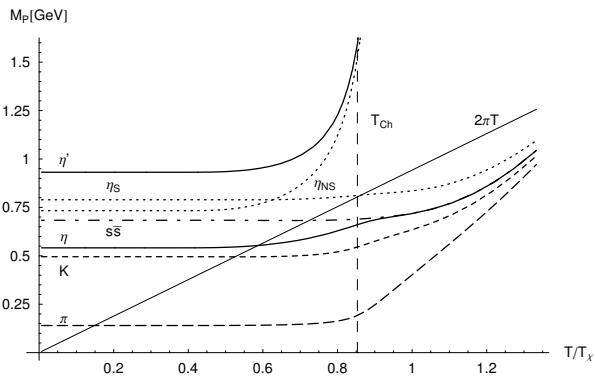}}
\caption{The $T/T_\chi$ dependence of the meson masses for 
$T_\chi = 1.17\, T_{\rm Ch}$.  The meaning of all symbols 
is again the same as in Figs. \ref{figPseudoTc85},
\ref{figPseudoTc107} and \ref{figPseudoTc128}.
The increase of $M_{\eta'}(T)$ is huge, as high as 10 GeV 
around $T=0.18$ GeV, since the masses in the 
$\eta$-$\eta'$ complex are calculated (as in Figs. \ref{figPseudoTc85},
\ref{figPseudoTc107} and \ref{figPseudoTc128}, 
where $T_\chi$ is, however, lower)  
with the pure Yang-Mills topological susceptibility 
\cite{Alles:1996nm}, which decreases comparatively slowly.
  }
\label{figPseudoTc150}
\end{figure}

The temperature behavior of the pseudoscalar meson masses for this
case is shown in Fig. \ref{figPseudoTc150}, and
it is clear from the understanding gained on the
previous choices of $T_\chi$.
The increase of the $\eta$ and $\eta'$ masses starts as early as
$T/T_\chi=0.6$, and, in the case of the $\eta'$ mass, it
is even more pronounced, the maximal $M_{\eta'}(T)$ being
some 10 GeV. Its fall towards the $s\bar s$ mass
(and the fall of the mixing angle from almost $90\deg$ to 0)
is not seen in Fig. \ref{figPseudoTc150} since the approach
to the asymptotic behavior happens at higher temperatures.

Let us summarize our findings from the WV relation used consistently 
with the pure Yang-Mills topological susceptibility: only very 
low, unrealistic values of $T_\chi$, below some 70\% of $T_{\rm Ch}$,
would yield the drop in the $\eta$ and $\eta'$ masses which would allow the 
enhancement \cite{Kapusta:1995ww,Huang:1995fc} of the $\eta$ and $\eta'$
production as in Pisarski-Wilczek scenario. For all higher 
values of $T_\chi$ we found mostly increasing $\eta$ and $\eta'$ masses
allowing no such enhancement in their production. 
The increase of the $\eta$ mass in Figs. \ref{figPseudoTc107}, 
\ref{figPseudoTc128}, and \ref{figPseudoTc150} is however moderate
and relatively slow, so that in our qualitative treatment we do 
not venture to make any statement about a possible suppression 
of the $\eta$ production either. 
In contrast to that, the enhancement of the $\eta'$ mass is so 
dramatic for $T_\chi \agt T_{\rm Ch}$, that one can 
argue that the $\eta'$ production must be ever more 
suppressed for these values of $T_\chi$ -- of course, provided that
the $\eta'$ mass increase observed in the present approach is at all 
genuine, and not an artefact.

\subsection{ Discussing caveats and 
probing other possibilities}
\label{caveatsPossib}

\begin{figure}[!tbp]
\centerline{\includegraphics[width=80mm,angle=0]{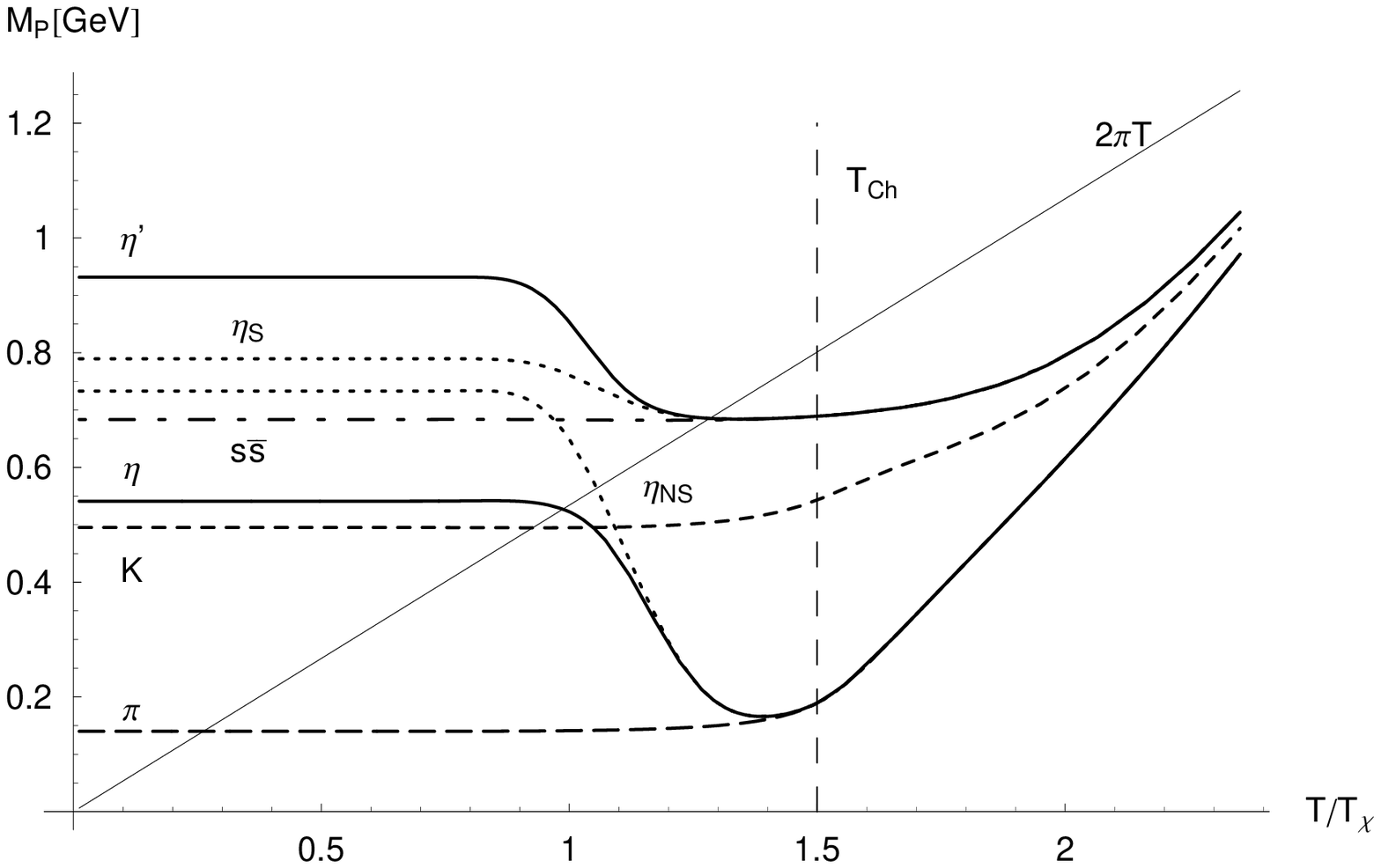}}
\centerline{\includegraphics[width=80mm,angle=0]{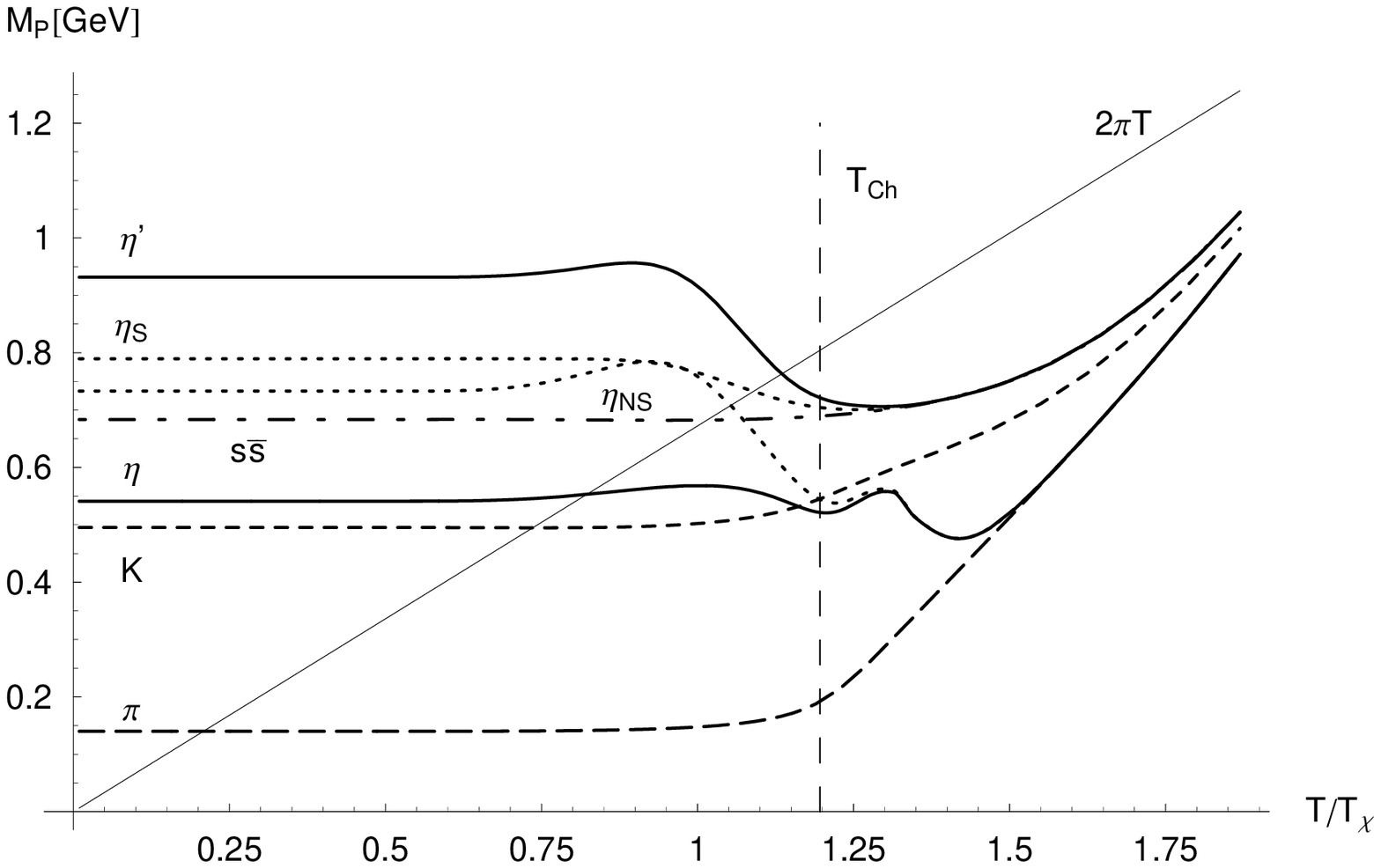}}
\centerline{\includegraphics[width=80mm,angle=0]{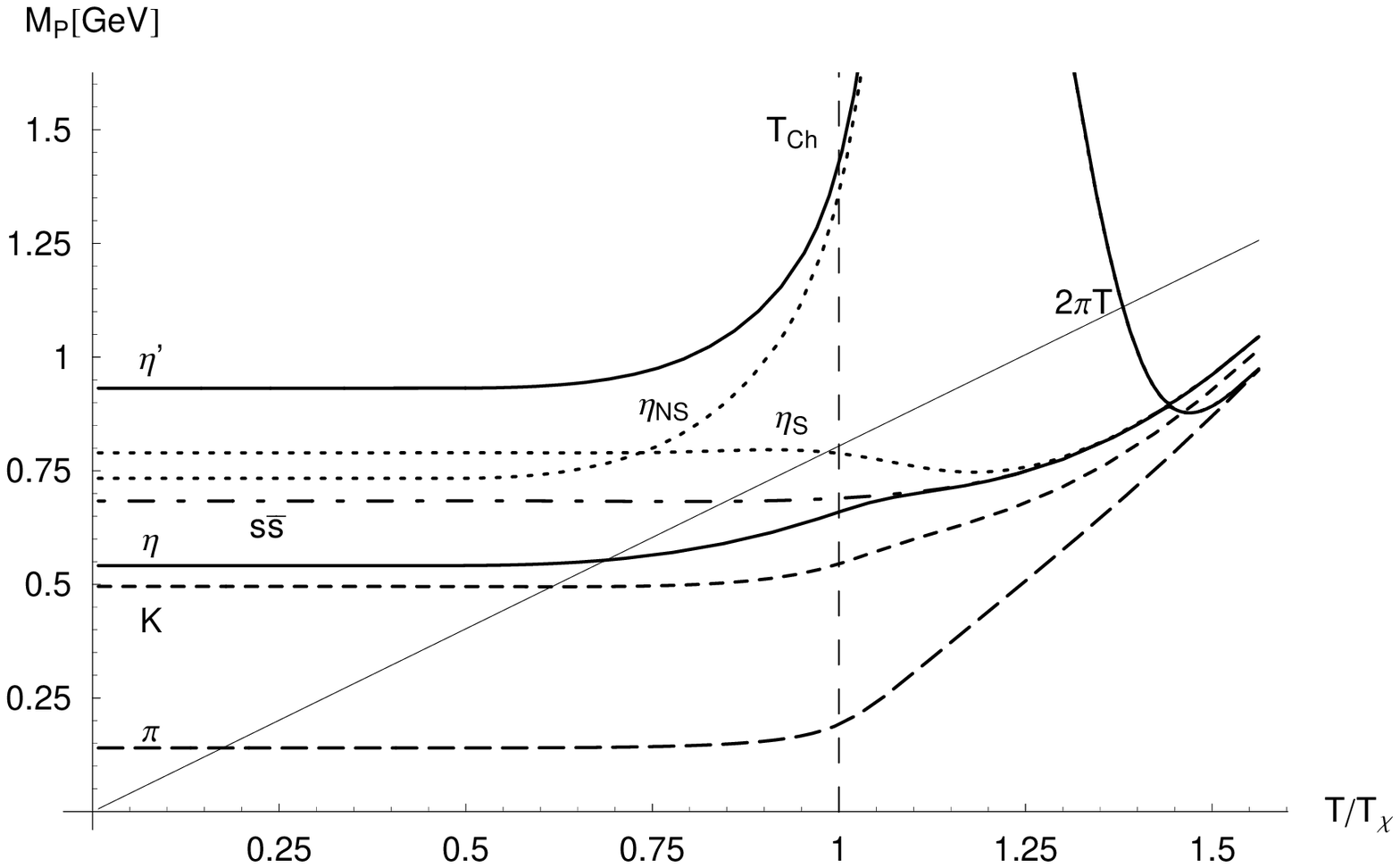}}
\centerline{\includegraphics[width=80mm,angle=0]{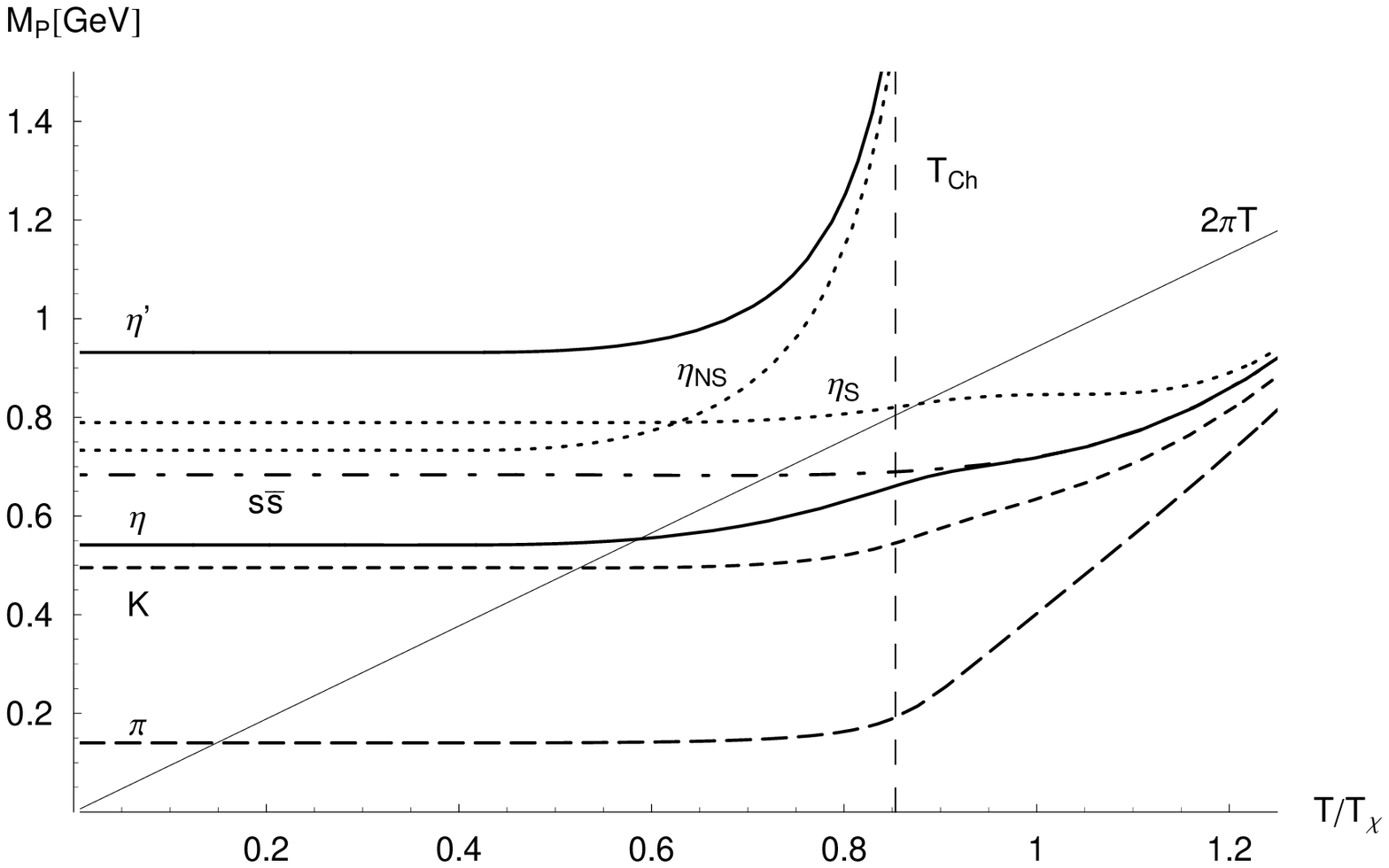}}
\caption{
The relative temperature dependence, on $T/T_\chi$, of 
the pseudoscalar meson masses for the SU(3) quenched 
topological susceptibility from the lattice \cite{Alles:1996nm}. 
{}From top to bottom, the diagrams 
correspond to $T_\chi = 0.667 \, T_{\rm Ch}$, 
$T_\chi = 0.836\, T_{\rm Ch}$, $T_\chi = T_{\rm Ch}$,
and $T_\chi = 1.17\, T_{\rm Ch}$.
That means the diagrams in this composite figure 
correspond to the individual
Figs. \ref{figPseudoTc85}, \ref{figPseudoTc107}, 
\ref{figPseudoTc128} and \ref{figPseudoTc150} (employing
the pure Yang-Mills topological susceptibility).
The description of the curves and symbols is 
the same as in these figures. 
In the third and fourth diagram, $M_{\eta'}$ exceeds the scale,
and the respective maxima are roughly 2.8 GeV and 7.7 GeV. 
What looks like a crossing of $M_{\eta}$ and $M_{\eta'}$ in
the third diagram, is really a very close anti-crossing, as 
clearly seen in the third diagram of Fig. \ref{figNf4QCD}.
}
\label{figSU3qQCD}
\end{figure}

While our bound-state approach, which includes both DChSB and
realistic {\it explicit} ChSB, makes it possible to use WV
relation, and the pion decay constant in it, for the finite
temperatures across the restorations of $U_A(1)$ and chiral
symmetries, there is no guarantee that the WV relation
does not ultimately fail at some $T$. Of course, it is not
likely that the WV relation, after being a good approximation 
at lower $T$, should fail almost immediately after $f_\pi$ 
starts falling with $T$. Thus, the behaviors of $\eta$ and 
$\eta'$ exhibited in the previous subsection should be 
considered at least as a serious qualitative indication
that, e.g., some $\eta'$ mass enhancement (and associated
suppression of $\eta'$ production) should exist at least 
over some $T$-interval, even though the WV relation may 
break down after certain $T$. At this point, nevertheless,
we must discuss and take into account the caveat that such
a drastic $\eta'$ mass increase (for $T_\chi \agt T_{\rm Ch}$)
suggest that it is in fact not a genuine effect, but at
least to a large extent an artefact of our approximations
and assumptions. 

One can think of the large $N_c$ approximation, in which
the WV relation (with $\chi = \chi_{\rm YM}$) was 
derived, as a possible cause of unreliability of 
the WV relation after some $T$. 
The avenue towards improvement and better insight can then be sought 
in the recent work by Shore \cite{Shore:2006mm,Shore:2007yn}, which 
contains what amounts to the generalization of the WV relation valid to 
{\it all orders} in $1/N_c$. Instead of the topological susceptibility
$\chi$, Shore's equations employ the full QCD topological charge 
parameter $A$ (but since it is at present not known, Shore himself 
\cite{Shore:2006mm,Shore:2007yn} had to approximate it by its 
lowest-order approximation in the $1/N_c$ expansion, namely by 
the pure Yang-Mills topological susceptibility $\chi_{\rm YM}$). 
Also, besides $f_\pi$, the kaon decay constant and four different 
$\eta$ and $\eta'$ decay constants appear in  
the pertinent Shore's equations (2.14)-(2.16) \cite{Shore:2006mm}.
Very illustrative is also Eq. (2.19) in Ref. \cite{Shore:2006mm}, 
which after the $1/N_c$ expansion gives back the WV relation 
(\ref{WittenVenez}) as the lowest-order approximation.  
So, instead of the WV relation, Shore's generalization might 
be used in our bound-state approach. Indeed, this research has 
been initiated and is in progress \cite{DGMORinProgress}, and 
there are already some preliminary results (notably at 
$T_\chi = T_{\rm Ch}$, but the presentation of this material 
is beyond the scope of the present paper.
Nevertheless, already 
on the basis of the insights from the present work we can 
and should make some comments on the indicative fact 
that the results from Shore's generalization at $T>0$ are 
rather close \cite{DGMORinProgress} to our present results from the WV relation
provided the same topological susceptibility $\chi$ is used 
instead of the full QCD topological charge parameter $A$.

It is instructive to compare the WV relation (\ref{WittenVenez}) 
with Shore's \cite{Shore:2006mm} Eq. (2.19).
The refinement brought by the presence of the five additional
decay constants in Shore's generalization (instead of just $f_\pi$, 
as in the WV relation) immediately suggests that the $\eta'$ mass 
increase would not be as drastic as the ones obtained from the 
original WV relation for $T_\chi/T_{\rm Ch} \agt 1$. Namely, these 
additional decay constants are all affected by the strange quarks;
the way $f_{s\bar s}$ contributes to the $\eta$ and $\eta'$ decay 
constants can be seen in, e.g., Ref. \cite{Feldmann:1999uf} or 
Appendix of Ref. \cite{Kekez:2000aw}. Thus, they diminish with $T$ 
slower than $f_\pi$,
as shown in Fig. \ref{figfp} for $f_K$ and $f_{s\bar s}$. 
In keeping with that, our preliminary results indeed show that the 
$\eta'$ mass increase is not as pronounced as in the present paper. 
However, the reduction is just about some 20\%, and, as already said, 
the results are on the whole rather similar. The refinement through 
the decay constants (other than $f_\pi$) thus improves, but does not 
really cure the suspicious $\eta'$-mass behavior. This indicates
that the probable main problem is the usage of the topological 
susceptibility 
$\chi=\chi_{\rm YM}$ instead of the presently unknown full QCD 
topological charge parameter $A$. While the usage of the pure
Yang-Mills topological susceptibility $\chi_{\rm YM}$ often 
turns out to be a reasonable approximation at $T=0$ (e.g., see
Refs. \cite{Klabucar:1997zi,Kekez:2000aw,Kekez:2001ph,Kekez:2005ie,
Feldmann:1999uf,Shore:2006mm,Shore:2007yn}),
its `slow' temperature dependence contributes crucially to the
perceived artefact of the present approach -- the blow-up of the 
$\eta'$ mass. This motivates us to use the related topological 
susceptibilities obtained on the lattice \cite{Alles:1996nm} in the 
presence of quarks, namely the SU(3) quenched and the four-flavor 
QCD case, as Fig. \ref{figFit3AllesTopSusc} shows that they `melt' 
faster. Thus, they hopefully provide a better approximation to 
the $T$-dependence of the full QCD topological charge parameter,
since we expect that, as usual, this quantity of the full QCD 
will be less resistant to the temperature-induced changes than 
the pure-glue, quarkless $\chi_{\rm YM}$.

\begin{figure}[!tbp]
\centerline{\includegraphics[width=80mm,angle=0]{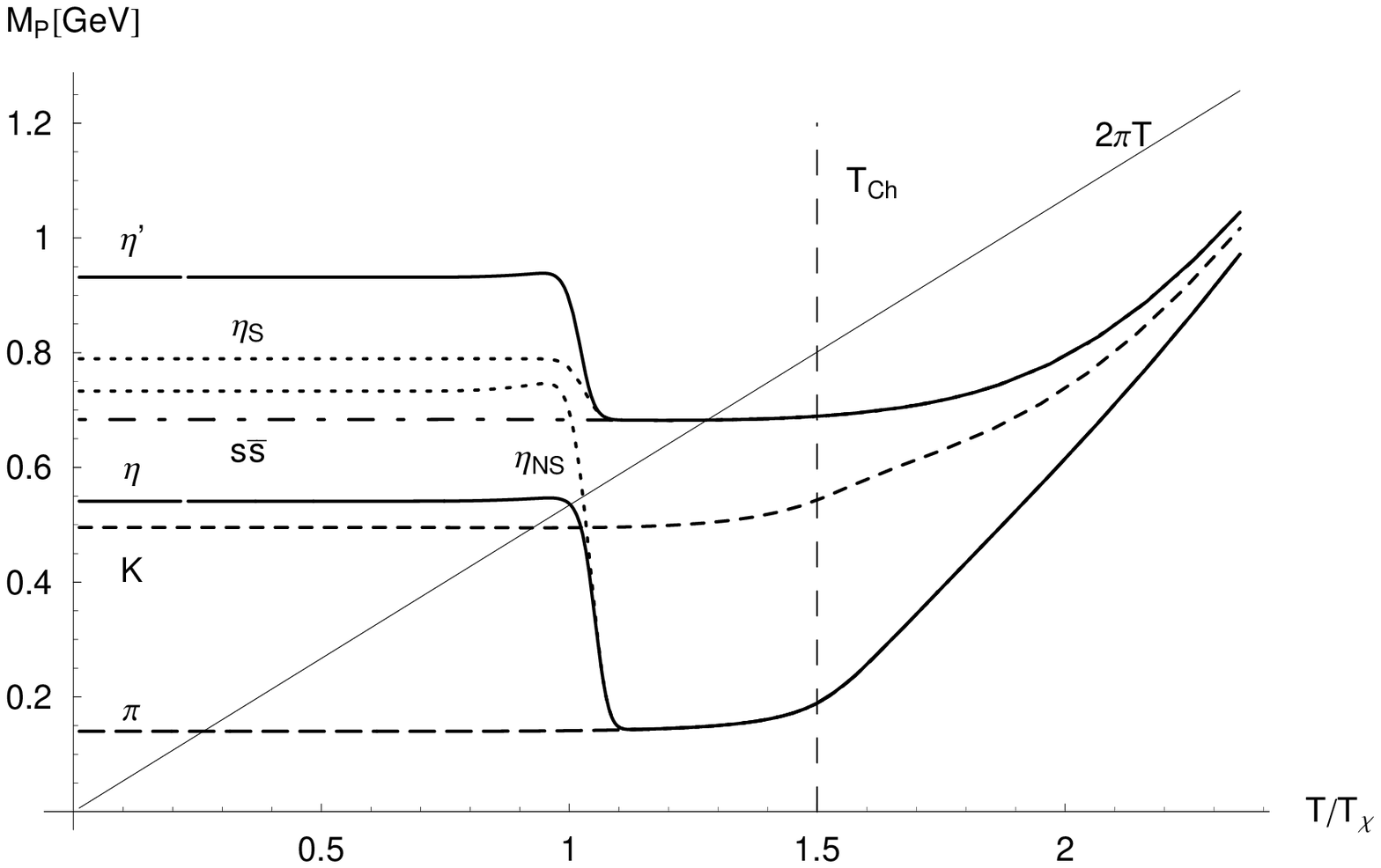}}
\centerline{\includegraphics[width=80mm,angle=0]{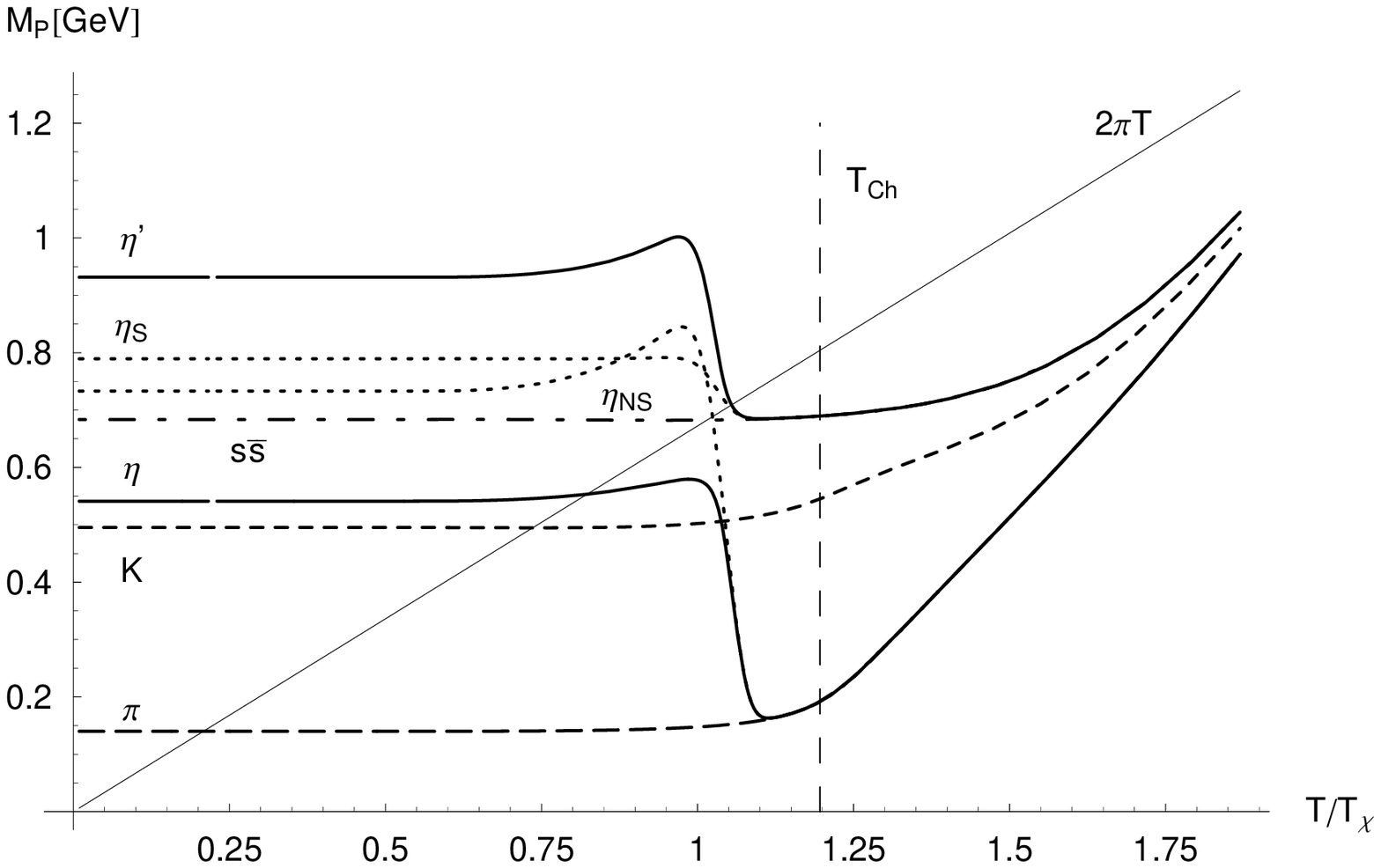}}
\centerline{\includegraphics[width=80mm,angle=0]{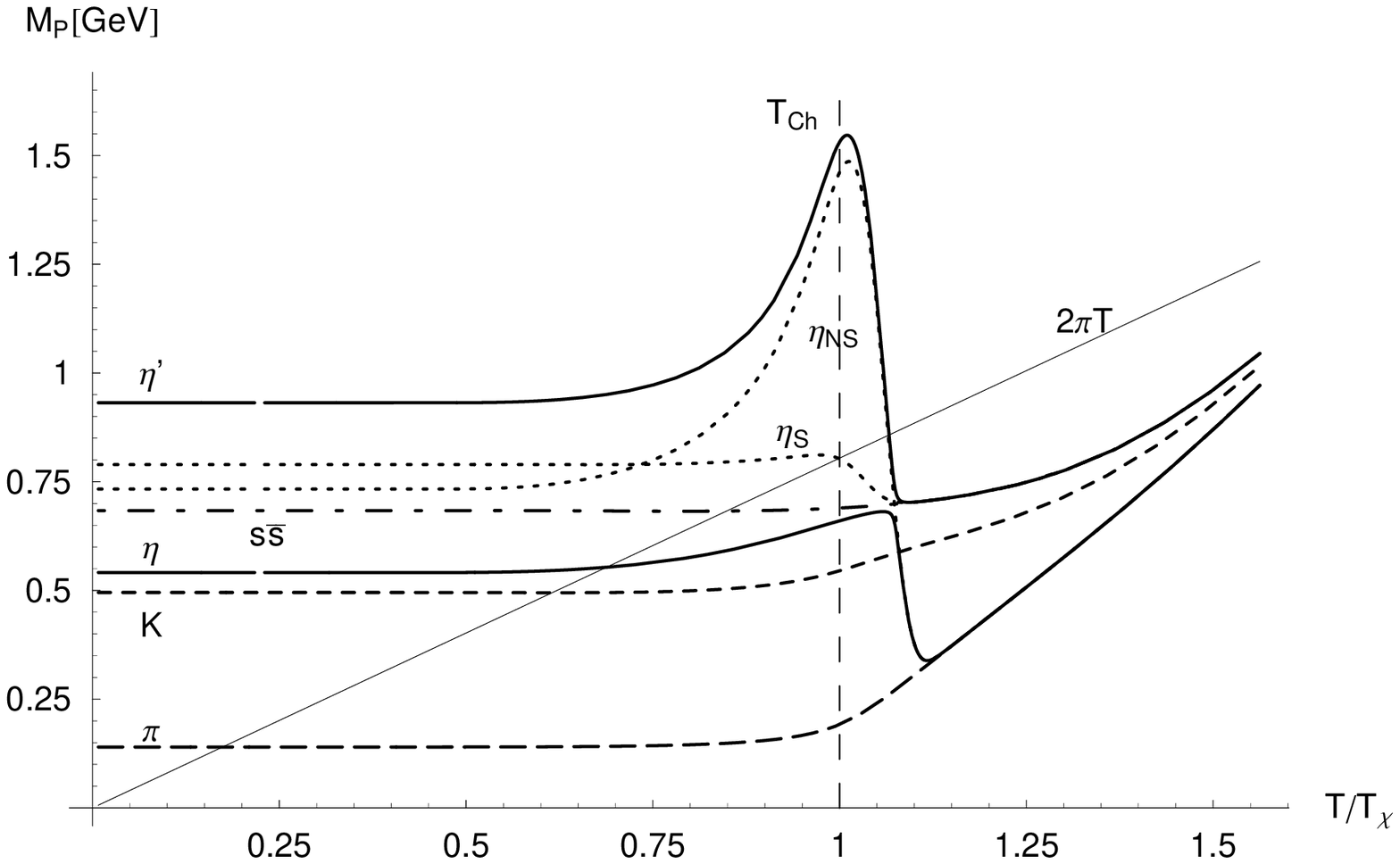}}
\centerline{\includegraphics[width=80mm,angle=0]{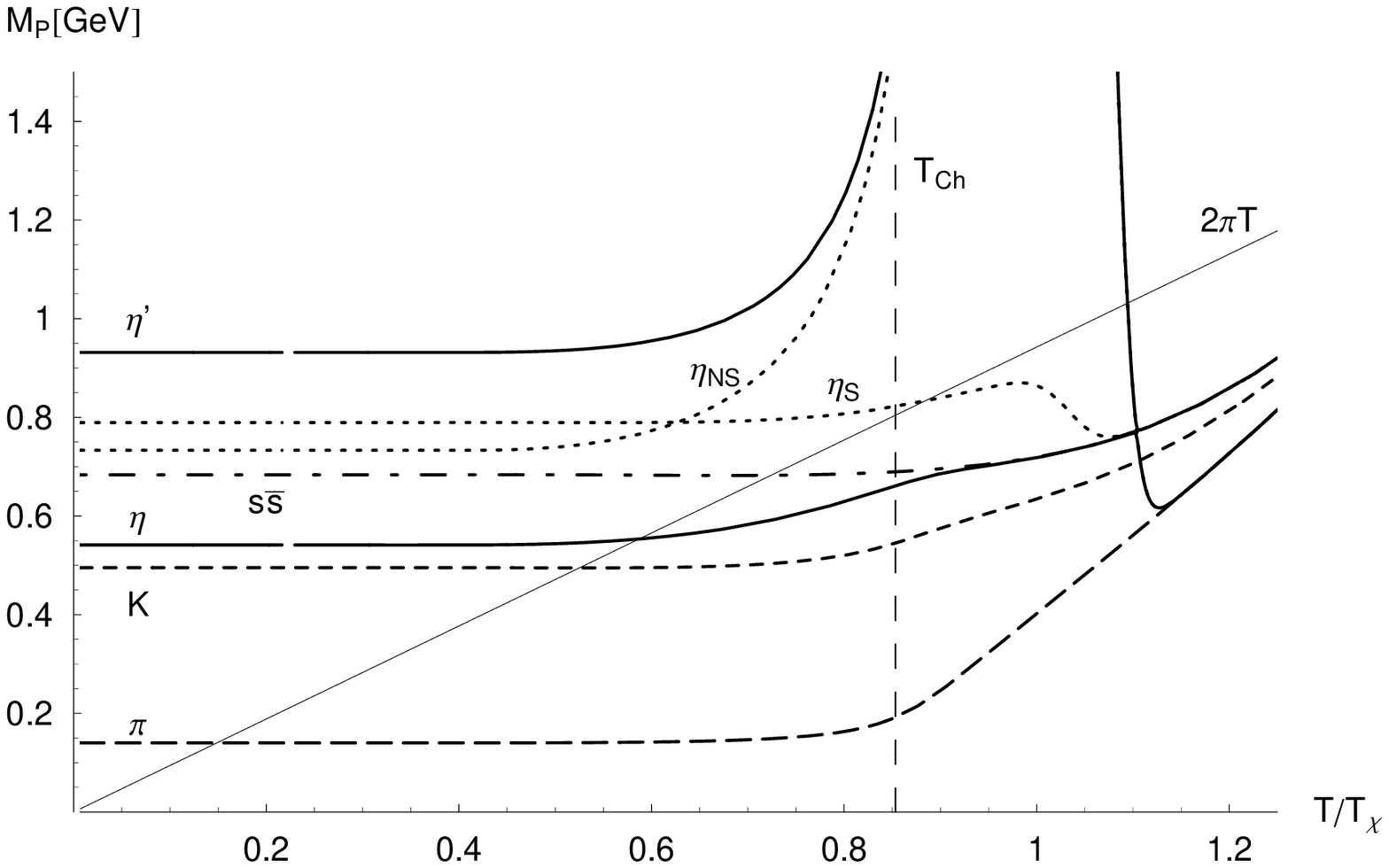}}
\caption{
The relative temperature dependence, on $T/T_\chi$,
of the pseudoscalar meson masses. The masses in the
$\eta$-$\eta'$ complex are calculated employing the  
topological susceptibility for the four-flavor QCD taken 
from the lattice. 
{}From top to bottom, the diagrams
correspond to $T_\chi = 0.667 \, T_{\rm Ch}$,
$T_\chi = 0.836\, T_{\rm Ch}$, $T_\chi = T_{\rm Ch}$,
and $T_\chi = 1.17\, T_{\rm Ch}$.
That is, the diagrams in this composite figure 
correspond to the parallel diagrams in the 
previous figure [obtained with the SU(3) quenched $\chi(T)$],
as well respective as the individual Figs. \ref{figPseudoTc85}, 
\ref{figPseudoTc107}, \ref{figPseudoTc128} and 
\ref{figPseudoTc150} (obtained with 
the pure Yang-Mills topological susceptibility).
The description of the curves and symbols is
the same as in these figures.
In the fourth diagram, $M_{\eta'}$ exceeds the scale,
and its maximum is 6.4 GeV. 
}
\label{figNf4QCD}
\end{figure}

In Fig. \ref{figSU3qQCD} we see the results for 
the SU(3) quenched QCD data,
and in Fig. \ref{figNf4QCD}, the results for the four-flavor QCD
data [as fitted by, respectively, the dashed and the dotted curve
in Fig. \ref{figFit3AllesTopSusc}, resulting from 
Eq. (\ref{fitAlles+alTopolSusc})]. 

In the both figures, the same four values of the relative
susceptibility melting temperatures $T_\chi/T_{\rm Ch}$ 
are used as in the figures in the preceeding subsection.
Thus, in both Figs. \ref{figSU3qQCD} and \ref{figNf4QCD},
the four successive graphs starting from the top, 
correspond to the four respective Figs. \ref{figPseudoTc85}, 
\ref{figPseudoTc107}, \ref{figPseudoTc128} and \ref{figPseudoTc150},
which gave the $T$-dependence for $\chi = \chi_{\rm YM}$.
It was described in detail in the 
previous subsection, and on the basis of this we can easily
understand the behavior of the masses in the new figures
\ref{figSU3qQCD} and \ref{figNf4QCD}. What happens is that for 
the SU(3) quenched QCD, and especially for the four-flavor QCD,
the topological susceptibility melts faster than in the 
pure Yang-Mills case, so that the patterns seen in the previous
subsection for the pure Yang-Mills case are (roughly) reproduced
at somewhat higher values of $T_\chi$ for the SU(3) quenched case,
and even higher values of $T_\chi$ for the four-flavor case.
Thus, the Pisarski-Wilczek scenario is realized in the upper 
two diagrams of Fig. \ref{figNf4QCD} illustrating the four-flavor 
case, and just in the first diagram of Fig. \ref{figSU3qQCD}, but in 
any case we observe $M_\eta(T)$ become degenerate with $M_\pi(T)$ fast, 
before the chiral restoration temperature $T_{\rm Ch}$, while the 
behavior observed previously in Fig. \ref{figPseudoTc85} would
now be still seen {\it well above} $T_\chi/T_{\rm Ch} = 2/3$.
The second diagram from the top of Fig. \ref{figSU3qQCD} 
illustrates nicely how the bump in $M_{\eta_{NS}}$ starts 
growing but then fails to develop since now $\chi(T)$ melts 
too fast at $T_\chi/T_{\rm Ch}=0.836$,
in contrast to the mass bump in Fig. \ref{figPseudoTc107}.
That bump, namely the strong enhancement of the $\eta'$ 
thermal mass, is now for the both susceptibilities, i.e.,
in both Figs. \ref{figSU3qQCD} and \ref{figNf4QCD}, 
encountered in their third diagram, where 
$T_\chi/T_{\rm Ch} = 1$. For the four-flavor QCD, with
its faster melting $\chi(T)$, the enhancement is only 
some 60\%, but for the SU(3) quenched QCD [with an 
intermediate rate of the $\chi(T)$ melting], the 
enhancement is drastic so that $M_{\eta'}$ and 
$M_{\eta_{NS}}$ mass bumps exceed the displayed
scale of the figure, the maximum being roughly
$M_{\eta'} \approx 2.8$ GeV. For the four-flavor QCD, 
such a mass bump happens in the last, 
``highest-$T_\chi/T_{\rm Ch}$" diagram of 
Fig. \ref{figNf4QCD}. The last diagram in the SU(3) 
quenched case, Fig. \ref{figSU3qQCD}, of
course exhibits even stronger $M_{\eta'}$ enhancement, 
although weaker than in Figs. \ref{figPseudoTc128} 
and \ref{figPseudoTc150} in the pure Yang-Mills case.

In summary, we tested the sensitivity of our results on 
variations of topological susceptibility, and found that 
they are sensitive to the detailed temperature behavior 
of $\chi(T)$. The changes 
are readily understood 
as the consequences of the faster-melting $\chi$'s.
The relative-temperature behaviors of the $\eta$ mass, both 
in this and the previous subsection, is consistent with the 
$M_{\eta}(T)$ behaviors recently found in the NJL model by 
Costa et al. \cite{Costa:2005cz} for various melting rates 
of $\chi$.     The enhancement of the $\eta'$ mass 
remained the most conspicuous feature for the realistic 
cases $T_\chi \geq T_{\rm Ch}$, but it was reduced 
somewhat by the faster-melting susceptibilities. Although 
most often still suspiciously large, the $M_{\eta'}(T)$ 
enhancement looks reasonable for the case of the 
four-flavor QCD with $T_\chi = T_{\rm Ch}$ (the third 
diagram in Fig.  \ref{figNf4QCD}), where one should keep 
in mind that it would be further reduced 
in Shore's generalization.

\section{Summary and conclusions}
\label{Summary}

The separable model of Ref. \cite{Blaschke:2000gd} provides the 
good description of the nonstrange light mesons at $T=0$ and $T>0$.
We extended it to the strange sector: the kaon phenomenology was 
successfully obtained, along with the results on the fictitious 
$s\bar s$ pseudoscalar meson, needed for the subsequent calculations
of the $\eta$-$\eta'$ complex. The $\eta$-$\eta'$ phenomenology 
at $T=0$ was then also successfully reproduced,
now in this specific dynamical model \cite{Blaschke:2000gd}, 
by following 
Refs. 
\cite{Klabucar:1997zi,Kekez:2000aw,Kekez:2001ph,Kekez:2005ie}.
Thereby the stage was set for the $T>0$ extension of
the DS approach to $\eta$ and $\eta'$ 
\cite{Klabucar:1997zi,Kekez:2000aw,Kekez:2001ph,Kekez:2005ie},
which is
the main goal of the present paper.

The WV relation (\ref{WittenVenez}) enables the DS approach
\cite{Klabucar:1997zi,Kekez:2000aw,Kekez:2001ph,Kekez:2005ie}
to determine the masses in the $\eta$-$\eta'$ complex in a 
parameter-free way, from the calculated non-anomalous pseudoscalar 
$q\bar q$ masses, the calculated pion decay constant $f_\pi$ 
(resulting from our $q\bar q$ bound-state solutions) and the 
lattice results on the topological susceptibility $\chi$,
be it at $T=0$ or $T>0$. 

The results for the thermal evolution of the masses in the 
$\eta$-$\eta'$ complex thus depend on the choice of the 
topological susceptibility $\chi(T)$;
in addition, the results depend even more on the relation
between the chiral restoration temperature $T_{\rm Ch}$ 
and $T_\chi$, the temperature of melting of the topological 
susceptibility. Let us recall the oldest considered 
scenario, namely the Pisarski-Wilczek one, where
the expected drop of the $\eta$ and $\eta'$ masses 
as $\chi(T)$ drops with $T$, led to the expectations 
of the enhancement of $\eta$ and $\eta'$ multiplicities with 
growing $T$ \cite{Kapusta:1995ww,Huang:1995fc}. 
The (relative) $T$-dependences of the masses consistent
with this scenario are realized in Fig. \ref{figPseudoTc85},
the first diagram of Fig. \ref{figSU3qQCD} and the first 
two diagrams of Fig. \ref{figNf4QCD}. However, this  
requires a rather low $T_\chi$, noticeably below $T_{\rm Ch}$.
This is unrealistic, 
as the values of $T_\chi$ lower than $T_{\rm Ch}$ 
are excluded by the lattice; e.g.,
Alles {\it et al.} have
(in their notation, $T_\chi \to T_c$)  
$T_\chi \approx 260$ MeV \cite{Alles:1996nm,Boyd:1996bx}.
They identify this $T_\chi$ with $T_{\rm Ch}$, but for
the value appropriate to the pure Yang-Mills case, which 
is higher than $T_{\rm Ch}$ in the presence of quarks.
Also, there are the analogous, newer lattice results 
\cite{Gattringer:2002mr}, with even higher $T_\chi \approx 300$ MeV.
In the both cases, $T_\chi$
is the melting temperature of the topological
susceptibility of the {\it pure gauge}, Yang-Mills
theory. It is thus natural that it is above the characteristic
temperatures of the {\it full} QCD, which are lowered
(with respect to Yang-Mills) by 
the presence of the quark degrees of freedom.
This may cause (in analogy with $T_{\rm Ch}$) 
that from the high values characterizing the pure Yang-Mills case, 
the physically relevant $T_\chi$ (e.g., $T_\chi$ appropriate to 
the full QCD topological charge parameter) 
gets lowered to $T_c \approx T_{\rm Ch}$ of the full QCD 
-- but {\it not} to still lower values 
($T_\chi < T_{\rm Ch}$) generally required to realize 
the Pisarski-Wilczek scenario in our model. 
 Thus, our model indicates that this scenario 
and the hot QCD matter signal 
(increased $\eta$ and $\eta'$ multiplicity) associated with it
\cite{Kapusta:1995ww,Huang:1995fc} is excluded -- apart
from a possible partial exception,  
the third diagram in Fig. \ref{figNf4QCD}. 
This is the only $T_\chi \geq T_{\rm Ch}$ case where we observe 
that $M_\eta(T)$ suffers a significant fall. After 
$T= 1.1\, T_{\rm Ch}$, it drops to roughly half of 
its $T=0$ value, i.e., to $M_\pi(T)$ which is not
yet excessively thermally enhanced. Being preceded by a
(modest) rise, this fall of $M_\eta(T)$ maybe cannot
provide an easily detectable increase of the $\eta$ production,
but this is closest our DS approach gets to the 
Pisarski-Wilczek scenario for an acceptable value 
of $T_\chi/T_{\rm Ch}$.

Having mentioned the full QCD topological charge parameter
appearing in Shore's generalization of the WV relation,
let us add that the present paper and issues risen therein, 
obviously provide additional motivation for lattice calculations
to determine this quantity and its melting temperature $T_\chi$.
We expect that this $T_\chi$ may turn out to be equal or 
close to the other characteristic temperatures of the full
QCD, in analogy with the result $T_c \approx T_{\rm Ch}$;
e.g., see
Refs. \cite{Karsch:2001cy,Gattringer:2002dv,Hatta:2003ga}.

This lowering of $T_\chi$ to $T_{\rm Ch}$ would help controlling
the $\eta'$ mass enhancements appearing for $T_\chi \agt T_{\rm Ch}$,
which become more and more dramatic as $T_\chi/T_{\rm Ch}$ grows. 
Except in the third diagram of Fig. \ref{figNf4QCD}, these 
enhancements are so drastic, that they must to a large extent
be an artefact of our approximations and assumptions, and one
cannot rely on them quantitatively. On the other hand, their 
persistent appearances are at least a qualitative indication 
that probably $M_{\eta'}(T\agt T_{\rm Ch}) > M_{\eta'}(0)$
in some $T$-interval around $T_{\rm Ch}$.
Thus, even though we cannot make quantitative predictions,
on the basis of our results we expect a  
suppression of the $\eta'$ yield after the onset of the 
chiral symmetry restoration in the hot QCD medium. 
We thus propose that the measurements of the $\eta'$
multiplicity should be undertaken at RHIC and LHC.
Even if the $\eta'$ suppression would not be seen,
it would still be an interesting result because it
would falsify the $T>0$ extension of the so far successful 
bound-state approach to $\eta$ and $\eta'$
\cite{Klabucar:1997zi,Kekez:2000aw,Kekez:2001ph,Kekez:2005ie}.

In contrast to $\eta'$, the mass of $\eta$ does not exhibit 
a marked rise in any figure (except the gradual approach to
the limit of two lowest Matsubara frequencies). In fact, 
beyond the chiral restoration 
temperature, it rises somewhat more slowly than the pion 
mass. On the basis of these masses, one cannot expect any 
suppression of the relative $\eta/\pi^0$ multiplicity.
This is in agreement with the recent experimental results
of PHENIX collaboration, which recently found a common
suppression pattern of $\eta$ and $\pi^0$ mesons at high 
transverse momentum in Au+Au collisions \cite{Adler:2006hu}. 

\vspace{5mm}

{\bf Acknowledgments} 

We thank D. Blaschke and Yu.L. Kalinovsky for useful discussions.
A.E.R.~acknowledges support by RFBR grant No. 05-02-16699, the
Heisenberg-Landau program and the HISS Dubna program of the Helmholtz
Association. D.H.~and~D.K. were supported by MZT project 
No.~119-0982930-1016. D.K. acknowledges the hospitality of 
Abdus Salam ICTP at Trieste.



\end{document}